\documentclass[12pt,a4paper]{article}
\begin{document}


\begin{titlepage}
\begin{flushright}
{KOBE-TH-00-08}\\
{ICRR-REPORT-467-2000-11}
\end{flushright}
\vspace{1cm}
\begin{center}
{\LARGE Vacuum  Structure of Twisted Scalar Field Theories}\\
\vspace{5mm}
{\LARGE on $M^{D-1}\otimes S^1$}

\vskip1truein
{\large Hisaki Hatanaka
\footnote{E-mail: {\tt hatanaka@icrr.u-tokyo.ac.jp}}
}\\
\vspace*{2mm}
{\it Institute for Cosmic Ray Research,\\
University of Tokyo, Kashiwanoha 5-1-5, Kashiwa 277-8582, Japan}\\
\vspace*{6mm}
 {\large Seiho Matsumoto
 \footnote{E-mail: {\tt matsumoto@oct.phys.sci.kobe-u.ac.jp}}}
{\rm and} {} {\large Katsuhiko Ohnishi
\footnote{E-mail: {\tt katuhiko@oct.phys.sci.kobe-u.ac.jp}}}\\
\vspace*{2mm}
 {\it Graduate School of Science and Technology,\\
Kobe University, Rokkodai, Nada, Kobe 657-8501, Japan}\\
\vspace*{6mm}
{\large Makoto Sakamoto
\footnote{E-mail: {\tt sakamoto@phys.sci.kobe-u.ac.jp}}}\\
\vspace*{2mm}
{\it Department of Physics,\\
Kobe University, Rokkodai, Nada, Kobe 657-8501, Japan }
\vskip0.7truein

\end{center}

\centerline{\bf Abstract}
\vskip0.1truein
\setlength{\baselineskip}{0.6cm}
Scalar field theories on $M^{D-1}\otimes S^1$,
which allow to impose twisted boundary conditions
for the $S^1$ direction,
are studied in detail,
and several novel features overlooked so far are revealed.
One of characteristic features is the appearance of 
critical radii of the circle $S^1$,
at which some of symmetries are broken/restored.
A phase transition can occur at the classical level or 
can be caused by quantum effects.
Radiative corrections can restore broken symmetries
or can break symmetries for small radius $R$.
A surprising feature is that the translational invariance for
the $S^1$ direction can spontaneously be broken.
A particular class of coordinate-dependent vacuum
configurations is clarified and the $O(N)$ $\phi^4$
model on $M^{D-1}\otimes S^1$ is extensively studied,
as an illustrative example.
\end{titlepage}

\newpage
\section{INTRODUCTION}\label{intoro}
\setlength{\baselineskip}{0.6cm}

In recent years,
there has been renewal of interest in higher
dimensional field theories with extra dimensions
\cite{LED,RandallSundrum,DvaliShifman,HatanakaInamiLim}.
A considerable number of ideas
and scenarios have been proposed,
and some of physics at low energies could profoundly
be understood from a viewpoint of extra dimensions.
Although the subject of extra dimensions is not new and
a lot of studies have been made on this subject,
theoretical understanding of field theories with extra dimensions
seems to be far from complete.

In this paper,
we study scalar field theories on $M^{D-1}\otimes S^1$ in detail
and report several interesting properties overlooked so far.
The parameter space of such theories is,
in general,
much wider than that of ordinary field theories 
on the Minkowski space-time,
and is spanned by twist parameters specifying boundary
conditions\cite{Isham,Hosotani},
in addition to parameters appearing in the actions.
Physical consequences caused by twisted boundary conditions
turn out to be unexpectedly rich and many of them have not
been uncovered so far.

One of characteristic features of such theories is the appearance
of critical radii of the compactified space,
at which some of symmetries are broken/restored%
\cite{FordYoshimura,Toms,Spallucci,O(N)1,O(N)2}.
A phase transition can occur at the classical level,
or can be caused by quantum effects.
Radiative corrections would become important when a compactification
scale becomes less than the inverse of a typical mass scale,
and then some of broken symmetries
could be restored for small compactification scales,
or conversely some of symmetries could be broken.
Another characteristic and perhaps surprising feature is
the spontaneous breakdown of the translational invariance
of compactified spaces\cite{translation}.
When some of scalar fields obey twisted boundary conditions,
we must be careful in finding the vacuum configuration
because coordinate-{\it dependent} configurations of 
twisted scalar fields could lower the total energy
than that of constant configurations.
Among other things,
a phenomenologically important observation is that
twisted boundary conditions can break supersymmetry
spontaneously\cite{SUSY}.
This is probably expected from the fact that the breakdown of the
translational invariance directly causes the supersymmetry breaking 
because translations and supersymmetry transformations are
related through the supersymmetry algebra.
This mechanism will give a new type of spontaneous supersymmetry
breaking mechanisms in connection with compactification.
It would be of great interest to search for realistic supersymmetric
models with this supersymmetry breaking mechanism,
though this subject will not be treated in this paper.

The paper is organized as follows:
In the next section,
a general discussion about scalar field theories on
$M^{D-1}\otimes S^1$
is given and a particular class of such theories
whose coordinate-dependent
vacuum configurations have a simple form is classified.
As an illustrative example,
the $O(N)$ $\phi^4$ model on $M^{D-1}\otimes S^1$ is
studied at the classical level in Sec.\ref{classical}
and at the  one-loop level in Sec.\ref{D=4} and \ref{D>4}, 
in detail.
Many interesting phenomena are found there.
In Sec.\ref{kaluza}, 
the model is reanalyzed with Kaluza-Klein modes from
a $(D-1)$-dimensional field theory point of view.
Sec.\ref{discussions}
is devoted to discussions.
In Appendix A, the vacuum configuration which minimizes a potential
is given.
In Appendix B, the one-loop mass corrections of the
$O(N)$ $\phi^4$ model on $M^{D-1}\otimes S^1$ are computed.

\section{A GENERAL DISCUSSION}\label{general}

In this section,
we shall discuss general features of scalar field theories 
on $M^{D-1}\otimes S^1$.
Let us consider an action which consists of $N$ real scalar fields
$\phi_i$ $(i=1,2,\cdots,N)$
	\begin{equation}
		S
	=
		\int d^{D-1}x
		\int^{2\pi R}_0 dy
			\left\{
				-\frac{1}{2}\sum_{i=1}^N
					\partial_A\phi_i(x^{\nu},y)
					\partial^A\phi_i(x^{\nu},y)
				-V(\phi)
			\right\},
	\label{action}
	\end{equation}
where the index $A$ runs from $0$ to $D-1$
with the metric
${\rm diag}(\eta_{AB}) =(-,+,+,\cdots ,+)$
and
$x^\nu$ $(\nu = 0,1,\cdots ,D-2)$
and $y$ are the coordinates of $M^{D-1}$ and $S^1$,
respectively.
The radius of the circle $S^1$ is denoted by $R$.
Suppose that the action has a symmetry $G$,
which must be a subgroup of $O(N)$.
Since $S^1$ is multiply-connected,
we can impose a twisted boundary condition on $\phi_i$
such as
	\begin{equation}
		\phi_i(x^\nu,y+2\pi R)
	=
		\sum_{j=1}^N
			U_{ij}\phi_j(x^\nu,y).
	\label{generalboundary}
	\end{equation}
The matrix $U$ must belong to $G$,
otherwise the action would not be single-valued.
If $U$ is not proportional to the identity matrix,
the symmetry group $G$ will be broken to its subgroup $H$,
which consists of all the elements of $G$ commuting with $U$,
{\it i.e.} 
\begin{equation}
H=\{h\ |\ hU=Uh, h\in G\}.
\end{equation}
Note that this symmetry breaking caused by the boundary condition
is not spontaneous but explicit.
In fact,
radiative corrections do not respect the symmetry $G$
but preserve only the symmetry $H$,
as we will see in Sec.\ref{D=4}. 

In order to discuss general properties of
the boundary condition (\ref{generalboundary}),
it is convenient to transform the matrix $U$ by means of 
an orthogonal transformation into the normal form.
This can be done by writing
$\phi_i$ as $Q_{ij}\phi_j^\prime$,
where $Q\in O(N)$.
The boundary condition (\ref{generalboundary}) can then
be replaced by
$\phi_i^\prime(x^\nu,y+2\pi R)=U^\prime_{ij}\phi^\prime_j(x^\nu,y)$
with $U^\prime=Q^{-1}UQ$.
It is known that any matrix $U$ belonging to $O(N)$ can be transformed,
by an orthogonal transformation,
into a block diagonal form whose diagonal elements are one of
$1$, $-1$, and a two dimensional rotation matrix\cite{Schutz}.
Then,
the block-diagonalized matrix $U^\prime$ may be written into the form
	\begin{equation}
		U^\prime
	=
		\left(
			\begin{array}{cc}
				\begin{array}{ccc}
					{\bf 1}_{L_0}	&		&	\\
                    &  {\bf 1}_{\scriptstyle \frac{L_{1}}{2}}
						\otimes r(\alpha_1)&  \\
					& &{\bf 1}_{\scriptstyle \frac{L_{2}}{2}}
						\otimes r(\alpha_2)
				\end{array}
				& 0 \\
				0&
				\begin{array}{ccc} 
					\ddots & & \\
					& {\bf 1}_{\scriptstyle \frac{L_{M-1}}{2}}
						\otimes r(\alpha_{M-1}) & \\
					& & -{\bf 1}_{L_M}
				\end{array}
			\end{array}
		\right),
		\label{bigmatrix}
	\end{equation}
where ${\bf 1}_L$ denotes the $L\times L$ unit matrix and
$r(\alpha)$ is a two dimensional rotation matrix defined by
	\begin{equation}
		r(\alpha)
	=
		\left(
			\begin{array}{cc}
				{\rm cos} (2\pi\alpha)	&	-{\rm sin} (2\pi\alpha) \\
				{\rm sin} (2\pi\alpha)	 &	{\rm cos} (2\pi\alpha)
			\end{array}
		\right) .
		\label{2dimrotation}
	\end{equation}
The numbers $L_l$ $(l=0,1,\cdots,M)$ satisfy
	\begin{equation}
		L_0+L_1+\cdots +L_M=N
	\label{sumofl}
	\end{equation}
and the rotation angles $\alpha_l$ are arranged as
	\begin{equation}
		0\equiv\alpha_0<\alpha_1<\alpha_2<
		\cdots
		<\alpha_{M-1}<\alpha_M\equiv\frac{1}{2}\ .
	\label{orderofalpha}
	\end{equation}
Redefining the fields in this way and dropping the primes,
we may rewrite the boundary condition (\ref{generalboundary})
into the following set of the boundary conditions:
	\begin{equation}
		\phi_{a_0}^{(\alpha_0)}(y+2\pi R)
	=
		+\phi_{a_0}^{(\alpha_0)}(y)
	\qquad
		{\rm for}
	\quad
		a_0=1,2,\cdots,L_0,
	\label{alpha0boundary}
	\end{equation}
	\begin{eqnarray}
		\ \ \ 
		\left(
			\begin{array}{c}
				\phi_{2b_k-1}^{(\alpha_k)}(y+2\pi R) \\
				\phi_{2b_k}^{(\alpha_k)}(y+2\pi R)
			\end{array}
		\right)
	&=&
		r(\alpha_k)
		\left(
			\begin{array}{c}
				\phi_{2b_k-1}^{(\alpha_k)}(y) \\
				\phi_{2b_k}^{(\alpha_k)}(y)
			\end{array}
		\right)
	\label{alphakboundary}
	\\
	&&
	\mbox{for $b_k=1,2,\cdots,\frac{L_k}{2}$ and $k=1,2,\cdots,M-1,$}
	\nonumber
	\end{eqnarray}
	\begin{equation}
		\phi_{a_M}^{(\alpha_M)}(y+2\pi R)
	=
		-\phi_{a_M}^{(\alpha_M)}(y)
	\qquad
	\mbox{for $a_M=1,2,\cdots,L_M$}\ .
	\label{alphamboundary}
	\end{equation}
Instead of the real fields
$\phi_{a_k}^{(\alpha_k)}$
$(k=1,2,\cdots,M-1)$,
we may sometimes use the complex fields
$\Phi_{b_k}^{(\alpha_k)}$
defined by
	\begin{eqnarray}
		\Phi_{b_k}^{(\alpha_k)}(y)
	&\equiv&
		\frac{1}{\sqrt{2}}
			\left(
				\phi_{2b_k-1}^{(\alpha_k)}(y)
				+
				i\phi_{2b_k}^{(\alpha_k)}(y)
			\right)
	\label{deflargephi}
	\\
	&&\mbox{%
	for $b_k=1,2,\cdots,\frac{L_k}{2}$ and $k=1,2,\cdots,M-1$,
	}
	\nonumber
	\end{eqnarray}
and then the boundary conditions (\ref{alphakboundary})
are simply rewritten as
	\begin{equation}
		\Phi^{(\alpha_k)}_{b_k}(y+2\pi R)
	=
		{\rm e}^{i2\pi \alpha_k}\Phi_{b_k}^{(\alpha_k)}(y).
	\label{PhiandE}
	\end{equation}
One might try to introduce the complex basis,
like Eq.(\ref{deflargephi}),
for the real fields
$\phi_{a_0}^{(\alpha_0)}$
and
$\phi_{a_M}^{(\alpha_M)}$.
However,
the numbers $L_{0}$ and $L_{M}$ are not
necessarily even integers,
so that it is possible to introduce the complex basis for
all the fields only when $L_0$ and $L_M$ are even integers.
($L_k$ are always even for $k=1,2,\cdots,M-1$.)

In the basis
(\ref{alpha0boundary})--(\ref{alphamboundary}),
it may be easy to see what is the unbroken symmetry $H$,
which is consistent with the boundary conditions
(\ref{alpha0boundary})--(\ref{alphamboundary}).
For instance,
if $G=O(N)$,
the boundary conditions
(\ref{alpha0boundary})--(\ref{alphamboundary})
turn out to break the symmetry $O(N)$ down to
	\begin{equation}
		H=O(L_0)\times
		U\left({\textstyle \frac{L_1}{2}}\right)\times\cdots
			 U\left({\textstyle \frac{L_{M-1}}{2}}\right)\times O(L_M).
	\label{groupH}		 
	\end{equation}
When $G$ is a subgroup of $O(N)$,
the symmetry $H$ may be given by a subgroup of Eq.(\ref{groupH}).
In this paper,
we will not try to classify the unbroken symmetries $H$ 
for general $G$,
although the classification will not be difficult.

In order to find the vacuum configuration of the fields,
one might try to minimize the potential $V(\phi)$.
This would,
however,
lead to wrong vacua in the present model\cite{translation}.
To find the true vacuum configuration,
it is important to take account of the kinetic term 
for the $S^1$ direction,
in addition to the potential term.
This is because the translational invariance for the
$S^1$ direction could be broken and
the vacuum configuration might be $y$-dependent.
Thus,
the vacuum configuration will be obtained by
solving a minimization problem of the following 
functional
\footnote{This is nothing but the potential in a
$(D-1)$-dimensional point of view.}:
	\begin{equation}
		{\cal E}[\phi,R]
			\equiv
		\int^{2\pi R}_0dy
			\left\{
				\frac{1}{2}\sum^N_{i=1}
				\left(\frac{d\phi_i(y)}{dy}\right)^2
				+
				V(\phi)
			\right\},
	\label{enedensity}
	\end{equation}
where we have assumed that the translational invariance
of the uncompactified $(D-1)$-dimensional Minkowski
space-time is unbroken.

To solve the minimization problem,
one might try to find configurations which are
the extrema of ${\cal E}[\phi,R]$,
{\it i.e.}
	\begin{equation}
		\frac{d^2\phi_i(y)}{dy^2}
	=
		\frac{\partial V(\phi)}{\partial\phi_i(y)}
	\qquad
	{\rm for}
	\quad	
		i=1,2,\cdots,N.
	\label{condition:of:minimum}
	\end{equation}
If we regard $y$ and $\phi_i$ as the time 
and the position of a particle in an $N$-dimensional space,
respectively,
then the differential equations (\ref{condition:of:minimum})
represent a motion of the particle
in the presence of the potential $-V(\phi)$,
subject to the constraints (\ref{generalboundary}) or 
(\ref{alpha0boundary})--(\ref{alphamboundary}).
In principle,
we could get the vacuum configuration by solving
the equations (\ref{condition:of:minimum}) 
with the boundary condition
(\ref{generalboundary}) and then by looking 
for a solution which gives
the minimum of ${\cal E}[\phi,R]$.
But in practice,
it would be hard to do so.

There is,
however,
a particular class of twisted boundary conditions for which we can
explicitly construct the vacuum configuration without fully
solving the equations (\ref{condition:of:minimum}).
Suppose that $G$ is a continuous symmetry and that
the twist matrix $U$ in Eq.(\ref{generalboundary}) is continuously
connected to the identity in $G$.
In other words,
there exists a continuous map $U(y)\in G$ from ${\bf 1}_N$ to
$U$ such that $U(0)={\bf 1}_N$ and $U(2\pi R)=U$.
For instance, the twist matrix $U$ in Eq.(\ref{bigmatrix})
can continuously be deformed into the identity matrix by
the following matrix $U(y)$:
\footnote{Since the matrix $U$ is assumed to
continuously be connected to the identity matrix,
$L_M$ (with $\alpha_M=1/2$) must be an even integer.}

	\begin{equation}
		U(y)
	=
		\left(
			\begin{array}{cc}
				\begin{array}{ccc}
					{\bf 1}_{L_0}	&		&	\\
                    &  {\bf 1}_{\scriptstyle \frac{L_{1}}{2}}
						\otimes r({\textstyle 
						\frac{\alpha_{\scriptscriptstyle 1}y}
						{2\pi R}})&  \\
					& &{\bf 1}_{\scriptstyle \frac{L_{2}}{2}}
						\otimes r({\textstyle 
						\frac{\alpha_{\scriptscriptstyle 2}y}
						{2\pi R}})
				\end{array}
				& 0 \\
				0&
				\begin{array}{cc} 
					\ddots &  \\
					& {\bf 1}_{\scriptstyle \frac{L_{M}}{2}}
						\otimes r({\textstyle 
						\frac{\alpha_{\scriptscriptstyle M}y}
						{2\pi R}})  \\
					& 
				\end{array}
			\end{array}
		\right),
		\label{bigmatrix2}
	\end{equation}
if the symmetry group $G$ contains 
	\begin{equation}
		\left(U(1)\right)^{\scriptstyle \frac{L_{1}}{2}}\times
		\left(U(1)\right)^{\scriptstyle \frac{L_{2}}{2}}\times
		\cdots
		\times
		 \left(U(1)\right)^{\scriptstyle \frac{L_{M}}{2}}
	\label{all:U(1)}
	\end{equation}
as a subgroup.
It is then convenient to introduce the new fields
$\bar{\phi}_i(y)$
$(i=1,2,\cdots,N)$
by
	\begin{equation}
		\phi_i(y)
	\equiv
		\sum_{j=1}^NU(y)_{ij}\bar{\phi}_j(y)
	\qquad
		{\rm for}
	\quad
		i=1,2,\cdots,N,
	\label{def:of:phibar}
	\end{equation}
or in the basis 
(\ref{alpha0boundary})--(\ref{alphamboundary})

	\begin{eqnarray}
		\phi_{a_0}^{(\alpha_0)}(y)
	&&\equiv\ \ 
		\bar{\phi}_{a_0}^{(\alpha_0)}(y)
	\qquad
	 	{\rm for}
	\quad
		a_0=1,2,\cdots,L_0,\nonumber\\
	\label{phibaralpha0}
		\left(
			\begin{array}{c}
				\phi_{2b_k-1}^{(\alpha_k)}(y) \\
				\phi_{2b_k}^{(\alpha_k)}(y)
			\end{array}
		\right)
	&&\equiv\ \ 
		r\left({\textstyle 
						\frac{\alpha_{k}y}{2\pi R}}\right)
		\left(
			\begin{array}{c}
				\bar{\phi}_{2b_k-1}^{(\alpha_k)}(y) \\
				\bar{\phi}_{2b_k}^{(\alpha_k)}(y)
			\end{array}
		\right)
	\label{phibaralphak}
	\\
	&&	\qquad {\rm for}
	\ \ 
		b_k=1,2,\cdots,{\textstyle \frac{L_k}{2}}
	\ \ 
		{\rm and}
	\ \ 
		k=1,2,\cdots,M.
	\nonumber
	\end{eqnarray}
Note that all the new fields $\bar{\phi}_i(y)$
$(i=1,2,\cdots,N)$ satisfy the periodic boundary condition.
Inserting Eqs.(\ref{phibaralphak})
into ${\cal E}[\phi,R]$,
we may write
	\begin{equation}
		{\cal E}[\phi,R]
	=
		{\cal E}^{(1)}[\bar{\phi},R]
		+
		{\cal E}^{(2)}[\bar{\phi},R],
	\label{dev:E}
	\end{equation}
where 
	\begin{eqnarray}
			{\cal E}^{(1)}[\bar{\phi},R]
	&\equiv&
			\int^{2\pi R}_0dy
				\left\{
					\sum_{l=0}^M\sum_{a_l=1}^{L_l}
					\frac{1}{2}
					\left(
						\frac{d\bar{\phi}_{a_l}^{(\alpha_l)}(y)}{dy}
					\right)^2 \right.
		\nonumber 
		\\
			& &-
			\left.
				\sum_{k=1}^M\sum_{b_k=1}^{L_{k/2}}
				\frac{\alpha_k}{R}
				\left(
					\frac{d\bar{\phi}_{2b_k-1}^{(\alpha_k)}(y)}{dy}
					\bar{\phi}_{2b_k}^{(\alpha_k)}(y)
					-
					\bar{\phi}_{2b_k-1}^{(\alpha_k)}(y)
					\frac{d\bar{\phi}_{2b_k}^{(\alpha)}(y)}{dy}
				\right)
			\right\}\ ,
			\label{def:E1}
		\\
		{\cal E}^{(2)}[\bar{\phi},R]
	&\equiv&
		\int^{2\pi R}_0dy
		\left\{
			\sum_{k=1}^M\sum_{a_k=1}^{L_k}
			\frac{1}{2}\left(\frac{\alpha_k}{R}\right)^2
			\left(\bar{\phi}^{(\alpha_k)}_{a_k}(y)\right)^2
			+
			V(\bar{\phi})
		\right\}\ .
	\label{def:E2}
	\end{eqnarray}
Here,
we have used the fact that $U(y)\in G$ for any $y$,
so that $V(\phi)=V(\bar{\phi})$.
Our strategy to find the vacuum configuration,
which minimizes the functional (\ref{dev:E}),
is as follows:
we shall first look for configurations
which minimize each of ${\cal E}^{(1)}[\bar{\phi},R]$ and
${\cal E}^{(2)}[\bar{\phi},R]$,
and then construct configurations which minimize both of
them simultaneously.
As discussed in Ref.\cite{O(N)1},
by expanding the fields $\bar{\phi}_i(y)$ in
the Fourier-series according to the periodic boundary
condition and by noting that $0\leq \alpha_l\leq 1/2$ for
$l=0,1,\cdots,M$,
it is easy to see that the minimum of ${\cal E}^{(1)}[\bar{\phi},R]$
can be realized by arbitrary real constants 
$\bar{\phi}_{a_l}^{(\alpha_l)}(y) = \bar{\phi}_{a_l}^{(\alpha_l)}$
($l=0,1,\cdots,M$ and $a_l=1,2,\cdots,L_l$).
Since ${\cal E}^{(2)}[\bar{\phi},R]$ includes no derivative
with respect to $y$,
any configurations minimizing the function
\footnote{Note that $\alpha_0=0$.}
	\begin{equation}
		\bar{V}(\bar{\phi})
	\equiv
		\sum_{l=0}^M\sum_{a_l=1}^{L_l}
		\frac{1}{2}\left(\frac{\alpha_l}{R}\right)^2
		\left(\bar{\phi}^{(\alpha_l)}_{a_l}\right)^2
		+
		V(\bar{\phi})
	\label{def:Vbar}
	\end{equation}
can give the minimum of the functional ${\cal E}^{(2)}[\bar{\phi},R]$.
Thus,
we conclude that
any constant configuration
$\bar{\phi}_{a_l}^{(\alpha_l)}$
which gives the minimum of the function $\bar{V}(\bar{\phi})$ can
minimize both of ${\cal E}^{(1)}[\bar{\phi},R]$ and
${\cal E}^{(2)}[\bar{\phi},R]$,
simultaneously
\footnote{
Precisely speaking,
for $\bar{\phi}^{(\alpha_M)}_{a_M}(y)$
(with $\alpha_M=1/2$),
we can take 
$\bar{\phi}^{(\alpha_M)}_{a_M}(y)
=
\bar{\phi}^{\prime (\alpha_M)}_{a_M}{\rm e}^{-i\frac{y}{2R}}$
with $\bar{\phi}^{\prime (\alpha_M)}_{a_M}$
being constants,
instead of $\bar{\phi}^{(\alpha_M)}_{a_M}(y)=
\bar{\phi}^{(\alpha_M)}_{a_M}$.
This choice is, however,
physically equivalent to the choice of 
$\bar{\phi}_{a_M}^{(\alpha_M)}(y)=\bar{\phi}_{a_M}^{(\alpha_M)}$.
}.
It follows that in terms of the original fields
the vacuum configuration can be taken to be of the form
	\begin{eqnarray}
		\langle \phi_{a_0}^{(\alpha_0)}(x^\nu,y)\rangle
	&=&
		\bar{\phi}_{a_0}^{(\alpha_0)}
	\qquad
	{\rm for}
	\quad
	a_0=1,2,\cdots,L_0,\nonumber\\
	\label{expect:phi0}	
		\left(
			\begin{array}{c}
				\langle\phi_{2b_k-1}^{(\alpha_k)}(x^\nu,y)\rangle \\
				\langle\phi_{2b_k}^{(\alpha_k)}(x^\nu,y)\rangle
			\end{array}
		\right)
	&=&
		r\left({\textstyle 
						\frac{\alpha_{k}y}{2\pi R}}\right)
		\left(
			\begin{array}{c}
				\bar{\phi}_{2b_k-1}^{(\alpha_k)} \\
				\bar{\phi}_{2b_k}^{(\alpha_k)}
			\end{array}
		\right)
	\label{expect:phik}
	\\
	&&	\quad{\rm for}
	\ \ 
		b_k=1,2,\cdots,{\textstyle \frac{L_k}{2}}
	\ \ 
		{\rm and}
	\ \ 
		k=1,2,\cdots,M,
	\nonumber
	\end{eqnarray}
or simply
	\begin{equation}
		\langle\phi_i(x^\nu,y)\rangle
	=
		\sum_{j=1}^NU(y)_{ij}\bar{\phi}_j,
	\label{expect:phi:and:U}
	\end{equation}
where $\bar{\phi}^{(\alpha_l)}_{a_l}$ or
$\bar{\phi}_j$ are taken to be the real
constants which give the minimum of the
function $\bar{V}(\bar{\phi})$ in Eq.(\ref{def:Vbar}).
Therefore,
we have found that the problem to find the vacuum configuration
simply reduces to the ordinary problem to minimize the
``potential" $\bar{V}(\bar{\phi})$ in a class of models
that the continuous map $U(y)$ in Eq.(\ref{bigmatrix2})
connects the identity matrix to the twist matrix $U$ 
in the group space $G$.
\footnote{
Some extensions of the above discussions will be found in
Ref.\cite{O(N)1}.
}

We would like to make two comments here.
The first comment is that in order to find the vacuum configuration
we must minimize the function $\bar{V}(\bar{\phi})$
in Eq.(\ref{def:Vbar}) but {\it not} the original potential $V(\phi)$.
The ``effective'' potential
$\bar{V}(\bar{\phi})$
includes an additional
mass term $(\alpha_l/R)^2(\bar{\phi}_{a_l}^{(\alpha_l)})^2/2$
for each field $\bar{\phi}^{(\alpha_l)}_{a_l}$.
This mass term turns out to become important in discussing the
symmetry breaking/restoration,
as we will see later.
The second comment is that if some of $\bar{\phi}^{(\alpha_k)}_{a_k}$
in Eqs.(\ref{expect:phik})
with $k\neq 0$ are non-vanishing,
the translational invariance under the transformations
$\phi_i(y)\to\phi_i(y+a)$ is spontaneously broken because
$\langle\phi_{a_k}^{(\alpha_k)}(y)\rangle$ become
$y$-dependent for non-vanishing $\bar{\phi}_{a_k}^{(\alpha_k)}$.
However,
the following modified translations still survive as a symmetry:
\begin{equation}
	\phi_i(y)\to 
	\sum_{j=1}^N\left(U(a)^{-1}\right)_{ij}\phi_j(y+a).
\end{equation}
This is because the above transformations preserve the vacuum
invariant.

\section{CLASSICAL ANALYSIS}\label{classical}

In the previous section,
we have discussed some general features of scalar field
theories on $M^{D-1}\otimes S^1$.
In the remaining sections,
we shall extensively study the $O(N)$ $\phi^4$ model whose
potential is given by
	\begin{equation}
		V(\phi)
	=
		\frac{m^2}{2}\sum_{i=1}^N(\phi_i)^2
		+
		\frac{\lambda}{8}
			\left(\sum_{i=1}^N(\phi_i)^2\right)^2,
	\label{pot:O(N):phi4}
	\end{equation}
as an illustrative example.
In this model,
the twist matrix $U$ in Eq.(\ref{generalboundary})
can be taken to be any element of $O(N)$.
The classical analysis of this model has been done
in Ref.\cite{O(N)1}.
Since the classical results will be used later,
we will briefly summarize the results below.

As discussed in the previous section,
any element of $O(N)$ can be transformed,
by means of an orthogonal transformation,
into the normal form (\ref{bigmatrix}).
In this basis,
the boundary condition (\ref{generalboundary})
reduces to Eqs.(\ref{alpha0boundary})--(\ref{alphamboundary})
and explicitly breaks the $O(N)$ symmetry down to
$H$ in Eq.(\ref{groupH}),
which is the subgroup of $O(N)$ commuting with
the twist matrix (\ref{bigmatrix}).

For $m^2>0$,
nothing happens at the classical level and the symmetry
$H$ remains unbroken in a whole range of $R$.
As we will see later,
this conclusion does not hold at the quantum level and
the spontaneous symmetry breakdown can occur in
some class of twisted boundary conditions.

The remaining analysis will be focused on the case of
$m^2\equiv -\mu^2<0$.
Let us first consider the model with $L_0\neq 0$.
Then,
there exist the fields $\phi_{a_0}^{(\alpha_0)}$
$(a_0=1,2,\cdots,L_0)$ which obey the periodic boundary condition.
The vacuum configuration turns out to be taken,
without loss of generality,
to be of the form
	\begin{equation}
		\langle
			\phi_1^{(\alpha_0)}(x^\nu,y)
		\rangle
	=
		\mu \sqrt{\frac{2}{\lambda}}
	\label{VE:of:phi1}
	\end{equation}
and other fields should vanish.
Thus,
the symmetry $H$ in Eq.(\ref{groupH}) is
spontaneously broken to
\footnote{
For $L_0=1$,
$O(1)$ means $Z_2$ and the $Z_2$ symmetry
broken to completely.
}
	\begin{equation}
		I=O(L_0-1)
		\times
		U\left({\textstyle \frac{L_1}{2}}\right)\times
		\cdots
		\times
		U\left({\textstyle \frac{L_{M-1}}{2}}\right)\times O(L_M),
	\label{groupI}
	\end{equation}
irrespective of the value of the radius $R$.

Let us next consider the case of $L_0=0$ and $L_1\neq 0$
($\alpha_1\neq 1/2$) with $N={\rm even}$.
Since $L_M$ (with $\alpha_M=1/2$) is even in this case,
the twist matrix $U$ is continuously connected to the identity matrix,
so that we can apply the arguments given in the previous section.
It follows that the problem to find the vacuum configuration
reduces to the problem to minimize the function
	\begin{equation}
		\bar{V}(\bar{\phi})
    =
		\sum_{k=1}^M\sum_{a_k=1}^{L_k}
		\frac{1}{2}
			\left[
				-\mu^2+\left(\frac{\alpha_k}{R}\right)^2
			\right]
		\left(\bar{\phi}_{a_k}^{(\alpha_k)}\right)^2
     +
	\frac{\lambda}{8}
		\left(
			\sum_{k=1}^M\sum_{a_k=1}^{L_k}
			\left(\bar{\phi}_{a_k}^{(\alpha_k)}\right)^2
		\right)^2
	\label{def:vbar}
	\end{equation}
with $\bar{\phi}_{a_k}^{(\alpha_k)}$ being constants.
As proved in Appendix A,
the configuration which minimizes $\bar{V}(\bar{\phi})$
is that for $R\leq \alpha_1/\mu$,
	\begin{equation}
			\bar{\phi}_{a_k}^{(\alpha_k)} = 0
			\quad\quad
			{\rm for}\ \ a_k=1,2,\cdots,L_k\ \ {\rm and}
            \ \ k=1,2,\cdots ,M
	\end{equation}
and that for $R > \alpha_1/\mu$,
     \begin{equation}
		\sum_{a_k=1}^{L_k}
		\left(\bar{\phi}_{a_k}^{(\alpha_k)}\right)^2
	  =
		\frac{2}{\lambda}
			\left[
					\mu^2-\left(\frac{\alpha_1}{R}\right)^2
			\right]
		\delta_{k,1}
	\qquad
		\mbox{for $k=1,2,\cdots,M$.}
    \end{equation}
Thus,
we find that the vacuum expectation values of
the original fields $\phi_{a_k}^{(\alpha_k)}$ can,
without loss of generality,
be taken into the form
	\begin{equation}
		\langle\phi_{a_1}^{(\alpha_1)}(x^\nu,y)\rangle
	=
		\left\{
		\begin{array}{lr}
			0
				& \mbox{for $R\leq\frac{\alpha_1}{\mu}$}
		\\
			\left(
				v\cos\left(\frac{\alpha_1y}{R}\right),
				v\sin\left(\frac{\alpha_1y}{R}\right),
				0,\cdots,0
			\right)
				& \mbox{for $R>\frac{\alpha_1}{\mu}$}
		\end{array}
		\right.,
	\label{ex:phi:and:R}
	\end{equation}
and other fields should vanish,
where $v=\sqrt{2(\mu^2-(\alpha_1/R)^2)/\lambda}$.
It follows that for $R\leq\alpha_1/\mu$ the symmetry $H$
with $L_0=0$ is unbroken,
while for $R>\alpha_1/\mu$ it is spontaneously broken to
	\begin{equation}
		I
	=
		U\left({\textstyle\frac{L_1}{2}-1}\right)
		\times
		U\left({\textstyle\frac{L_2}{2}}\right)
		\times
		\cdots
		\times
		U\left({\textstyle\frac{L_{M-1}}{2}}\right)
		\times
		O(L_M).
	\label{groupI:no:periodic}
	\end{equation}
If $L_1=L_2=\cdots=L_{M-1}=0$,
{\it i.e.}
$L_M=N$,
the vacuum configuration is still given by the form
(\ref{ex:phi:and:R}) but $\alpha_1$ and $a_1$
should be replaced by $\alpha_M(=1/2)$ and $a_M$,
respectively.
Further,
the symmetry $H=O(L_M)=O(N)$ is unbroken for
$R\leq 1/(2\mu)$ but is spontaneously broken to
	\begin{equation}
		I=O(L_M-2)=O(N-2)
	\label{groupI:O(N-2)}
	\end{equation}
for $R>1/(2\mu)$.
It is interesting to contrast this result with that
of the model with $L_0\neq 0$,
for which the symmetry $O(L_0)$ is spontaneously
broken to $O(L_0-1)$ irrespective of $R$.

Let us finally investigate the case of $L_0=0$ with
$N={\rm odd}$.
In this case,
we cannot apply the same method, as was done above,
to find the vacuum configuration because the twist matrix
$U$ is not continuously connected to the identity matrix
due to the fact that $\det U=-1$.
Nevertheless, we can show that the problem to find the
vacuum configuration for odd $N$ reduces to that for even $N$
(expect for $N=1$).
The trick is to add an additional real field $\phi_{N+1}(y)$
satisfying the antiperiodic boundary condition to the
action in order to form the $O(N+1)$ $\phi^4$ model.
The analysis given in Ref.\cite{O(N)1} shows that the
vacuum configuration for odd $N$ is exactly the same form
as Eq.(\ref{ex:phi:and:R}).
The exception is the model with $N=1$.
In this case,
there is no continuous symmetry and the $O(1)$ model has
only a discrete symmetry,
{\it i.e.} $G=H=Z_2$.
The $O(1)$ $\phi^4$ model has been investigated in
Ref.\cite{translation} and the vacuum configuration has been
found to be
	\begin{equation}
		\langle\phi(x^\nu,y)\rangle
	=
		\left\{
		\begin{array}{lr}
			0
				& \quad \mbox{for $R\leq\frac{1}{2\mu}$}
		\\
			\frac{2k\mu}{\sqrt{\lambda(1+k^2)}}
			{\rm sn}
			\left(
				\frac{\mu y}{\sqrt{1+k^2}},k
			\right)
				& \quad \mbox{for $R>\frac{1}{2\mu}$}
		\end{array}
		\right..
	\label{Z2:ex:phi:and:R}
	\end{equation}
Here,
${\rm sn}(u,k)$
is the Jacobi elliptic function whose period is
$4K(k)$,
where $K(k)$ denotes the complete elliptic function of the first kind.
The parameter $k$ ($0\leq k<1$) is determined by the relation
$\pi R \mu =\sqrt{1+k^2}K(k)$.
Thus,
the $Z_2$ symmetry is unbroken for $R\leq 1/(2\mu)$,
while it is broken spontaneously for $R>1/(2\mu)$.

Before closing this section,
it may be instructive to give an intuitive
explanation why the symmetry restoration occurs for small
radius $R$ in the model with $L_0=0$ and $m^2<0$.
We first note that since $\phi_{a_k}^{(\alpha_k)}(y)$
($k\neq 0$) obeys the twisted boundary condition,
a non-vanishing vacuum expectation value of
$\phi_{a_k}^{(\alpha_k)}(y)$
immediately implies that it is $y$-{\it dependent},
otherwise it would not satisfy the desired boundary condition.
The $y$-dependent configuration of
$\langle\phi_{a_k}^{(\alpha_k)}(y)\rangle$
will induce the kinetic energy proportional to $1/R^2$.
Then, for large radius $R$,
non-vanishing
$\langle\phi_{a_k}^{(\alpha_k)}(y)\rangle$
for some $k$ are preferable because the origin is
not the minimum of the potential for $m^2<0$ and
because the contribution from the kinetic energy is
expected to be small.
Therefore,
for large radius $R$,
the symmetry $H$ and also the translational invariance
of $S^1$ will spontaneously be broken.
On the other hand,
for small radius $R$,
the contribution from the kinetic energy becomes large,
so that the $y$-independent configuration of 
$\langle\phi_{a_k}^{(\alpha_k)}\rangle$
is preferable and this implies that
$\langle\phi_{a_k}^{(\alpha_k)}\rangle$
should vanish.

\section{QUANTUM EFFECTS IN $D=4$}\label{D=4}

In the previous section,
we have investigated the vacuum structure of the
$O(N)$ $\phi^4$ model on $M^{D-1}\otimes S^1$
at the classical level.
In the following,
we shall take quantum corrections into account and
reanalyze the model at one-loop order.
We are,
in particular,
interested in the $R$-dependent part of mass corrections
and show how quantum effects alter the classical results.
Since asymptotic behavior of quantum corrections as
$R\to 0$ depends on the space-time dimension $D$,
we will discuss the case of $D=4$ in this section and $D>4$
in the next section.

We have learned from the previous two sections that the problem
to find the vacuum configuration of the model will reduce to the
problem to minimize the ``effective'' potential
\footnote{
We will discuss the effect of vertex corrections at
the end of this section.
}
	\begin{equation}
		\bar{V}(\bar{\phi})
	=
		\frac{1}{2}\sum_{l=0}^M\sum^{L_l}_{a_l=1}
			{\cal M}^2(\alpha_l,R)
			\left(\bar{\phi}^{(\alpha_l)}_{a_l}\right)^2
	+
		\frac{\lambda}{8}
			\left(
				\sum_{l=0}^M\sum_{a_l=1}^{L_l}
					\left(
						\bar{\phi}^{(\alpha_l)}_{a_l}
					\right)^2
			\right)^2,
	\label{vbar:quantum}
	\end{equation}
where the one-loop mass corrections have been taken into account
\footnote{
The wave function renormalization could modify the coefficient of the
second term on the right-hand side of Eq.(\ref{quantum:full:mass}),
although there is no such correction at one-loop order
in the present model.
}, {\it i.e.}
	\begin{eqnarray}
		{\cal M}^2(\alpha_l,R)
	&=&
		m^2+\left(\frac{\alpha_l}{R}\right)^2
		+
		\Delta m^2(\alpha_l,R),
	\label{quantum:full:mass}
	\\
		\Delta m^2(\alpha_l,R)
	&=&
		\frac{\lambda}{2}
		\sum_{m=0}^M
		(L_m+2\delta_{m,l})
		\zeta^{D=4}_{\rm ren}(\alpha_m,R).
	\label{quantum:mass}
	\end{eqnarray}
The $\Delta m^2(\alpha_l,R)$ denotes the one-loop mass
correction to the field $\phi_{a_l}^{(\alpha_l)}$ and
$\zeta^{D=4}_{\rm ren}(\alpha,R)$ corresponds to the contribution
from the one-loop self-energy diagram depicted in Fig.1.
\begin{figure}
 \begin{center}
 \setlength\unitlength{1mm}
  \begin{picture}(55,26)
  \put(10,5){\line(1,0){35}}
  \put(27.5,11.5){\circle{13}}
  \put(35,13.5){\makebox(0,0)[lb]{$\phi^{(\alpha)}$}}
  \end{picture}
 \end{center}
\caption{{\footnotesize
A one-loop self-energy diagram.
The field $\phi^{(\alpha)}$ propagating through the internal
line denotes one of the fields 
$\{ \phi^{(\alpha_0)}_{a_0},
\Phi^{(\alpha_k)}_{b_k},
\phi^{(\alpha_M)}_{a_M} \}$
in the basis of Eqs.
(\ref{alpha0boundary}), (\ref{deflargephi})
and (\ref{alphamboundary}).
}}
\label{mass}
\end{figure}
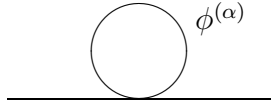
The function $\zeta_{\rm ren}^{D=4}(\alpha,R)$ is
computed in Appendix B
and is given by
	\begin{equation}
		\zeta^{D=4}_{\rm ren}(\alpha,R)
	=
		\frac{m}{4\pi^3R}
		\sum_{n=1}^\infty
		\frac{\cos(2n\pi\alpha)K_1(2n\pi Rm)}{n}
		,
	\label{def:reg:zeta}
	\end{equation}
where $K_\nu(z)$ denotes the modified Bessel function.
It is obvious from Eq.(\ref{quantum:mass}) that 
quantum corrections do not respect the $O(N)$ symmetry,
as mentioned previously.
This is due to the fact that the boundary condition 
(\ref{generalboundary})
or (\ref{alpha0boundary})--(\ref{alphamboundary})
explicitly breaks the $O(N)$ symmetry to $H$ 
in Eq.(\ref{groupH}) and
that the propagators of $\phi_{a_l}^{(\alpha_l)}$ depend on the
twist parameters $\alpha_l$ (see Eq.(\ref{propagator})).

For large radius $R\gg 1/m$,
Eq.(\ref{def:reg:zeta}) reduces to
	\begin{equation}
		\zeta^{D=4}_{\rm ren}(\alpha,R)
	\sim
		\frac{\cos(2\pi\alpha)m^{1/2}}{(2\pi)^3R^{3/2}}
		{\rm e}^{-2\pi Rm}.
	\label{zeta:for:large:R}
	\end{equation}
This implies that the quantum corrections are 
exponentially suppressed
and the classical analysis can be relied on for large radius
$R\gg 1/m$.
On the other hand,
for small radius $R\ll 1/m$,
Eq.(\ref{def:reg:zeta}) reduces to
	\begin{equation}
		\zeta^{D=4}_{\rm ren}(\alpha,R)
	\sim
		\frac{1-6\alpha+6\alpha^2}{48\pi^2R^2}.
	\label{zeta:for:small:R}
	\end{equation}
The above expression correctly reproduces the previously
known results for $\alpha=0$ (periodic) and
$\alpha=1/2$ (antiperiodic) given
in Refs.\cite{FordYoshimura,Toms,Spallucci}.
Since the mass corrections become large for $R\ll 1/m$,
we expect that quantum effects could alter the classical phase
structure for small $R$.
It is interesting to note that
the right-hand side of Eq.(\ref{zeta:for:small:R})
becomes negative if
	\begin{equation}
		\frac{3-\sqrt{3}}{6}
		<
		\alpha
		\leq
		\frac{1}{2}\ .
	\label{conditon:alpha:negative}
	\end{equation}
This fact suggests that quantum corrections could induce the
spontaneous symmetry breakdown,
as well as the restoration,
for small $R$.
Indeed,
we will see such an example later.

Let ${\cal M}^2(\alpha_P,R)$ be the lowest value between
${\cal M}^2(\alpha_l,R)$
($l=0,1,\cdots,M$),
that is,
${\cal M}^2(\alpha_P,R)$
$<$ ${\cal M}^2(\alpha_l,R)$ for $l\neq P$.
Then, the analysis given in Appendix A shows that
if ${\cal M}^2(\alpha_P,R)>0$,
all the vacuum expectation values of the fields turn out to vanish,
so that no spontaneous symmetry breaking occurs.
If ${\cal M}^2(\alpha_P,R)<0$,
some of the fields $\bar{\phi}^{(\alpha_P)}_{a_P}$ acquire
non-vanishing vacuum expectation values according to the relation
	\begin{equation}
		\sum^{L_P}_{a_P=1}
			\langle \bar{\phi}_{a_P}^{(\alpha_P)}\rangle^2
	=
		-\frac{2}{\lambda}{\cal M}^2(\alpha_P,R),
	\label{quantum:vev:of:phibar}
	\end{equation}
and other vacuum expectation values should vanish.
Thus,
the spontaneous symmetry breaking occurs.
If $P=0$ ({\it i.e.} $\alpha_P=0$),
the $O(L_0)$ symmetry in $H$ is broken to $O(L_0-1)$.
If $P\neq 0$, $M$
({\it i.e.} $\alpha_P\neq 0$, $1/2$),
the $U(L_P/2)$ symmetry is broken to $U(L_P/2-1)$.
If $P=M$ ({\it i.e.} $\alpha_P=1/2$),
the $O(L_M)$ symmetry is broken to $O(L_M-2)$.
Therefore,
for our purpose,
it is important to know the relative magnitudes of 
${\cal M}^2(\alpha_l,R)$'s
and the sign of the lowest ${\cal M}^2(\alpha_P,R)$.

Let us first investigate large radius behavior of 
${\cal M}^2(\alpha_l,R)$.
As was shown before,
the quantum corrections are exponentially suppressed for large $R$,
so that ${\cal M}^2(\alpha_l,R)$ will approximately 
be given by the classical
values,
{\it i.e.}
	\begin{equation}
		{\cal M}^2(\alpha_l,R)
	\sim
		m^2+\left(\frac{\alpha_l}{R}\right)^2
	\qquad
	\mbox{for $l=0,1,\cdots,M$}.
	\label{M:of:large:R}
	\end{equation}
Thus,
we have found the following increasing sequence of 
${\cal M}^2(\alpha_l,R)$
for large $R$:
	\begin{equation}
		{\cal M}^2(\alpha_0,R)<{\cal M}^2(\alpha_1,R)<
		\cdots
		<{\cal M}^2(\alpha_M,R).
	\label{order:of:M:large:R}
	\end{equation}
Now,
the analysis of the phase structure reduces to the classical one.
We will not repeat it here.

Let us next investigate small radius behavior of 
${\cal M}^2(\alpha_l,R)$.
For small $R$,
${\cal M}^2(\alpha_l,R)$ will reduce to
	\begin{equation}
		{\cal M}^2(\alpha_l,R)
	\sim
		m^2+\frac{C(\alpha_l)}{R^2},
	\label{M:of:small:R}
	\end{equation}
where
	\begin{equation}
		C(\alpha_l)
	=
		\alpha_l^2
		+
		\frac{\lambda}{96\pi^2}
		\sum^M_{m=0}(L_m+2\delta_{m,l})
		(1-6\alpha_m+6\alpha_m^2).
	\label{def:of:C}
	\end{equation}
The difference between ${\cal M}^2(\alpha_l,R)$ and 
${\cal M}^2(\alpha_m,R)$
is given by
	\begin{equation}
		{\cal M}^2(\alpha_l,R)-{\cal M}^2(\alpha_m,R)
	\sim
		\frac{f(\alpha_l)-f(\alpha_m)}{R^2},
	\label{subtract:f}
	\end{equation}
where
	\begin{equation}
		f(\alpha)
	\equiv
		\left(1+\frac{\lambda}{8\pi^2}\right)\alpha^2
		-
		\frac{\lambda}{8\pi^2}\alpha.
	\label{def:f}
	\end{equation}
It follows that if
	\begin{equation}
		\alpha_l
	>
		\frac{\lambda/(8\pi^2)}{1+\lambda/(8\pi^2)}
	\equiv
	    \alpha^\ast
    \qquad
	\mbox{for $l=1,2,\cdots,M$},
	\end{equation}
then we have an increasing sequence of ${\cal M}^2(\alpha_l,R)$,
{\it i.e.}
	\begin{equation}
		{\cal M}^2(\alpha_0,R)<{\cal M}^2(\alpha_1,R)<\cdots
		<{\cal M}^2(\alpha_M,R)\ .
	\label{order:M:small:alpha}
	\end{equation}
The above sequence immediately tells us that
if ${\cal M}^2(\alpha_0,R)>0$ with $L_0\neq 0$,
there is no spontaneous symmetry breaking,
while if ${\cal M}^2(\alpha_0,R)<0$,
the $O(L_0)$ symmetry in $H$ is spontaneously broken
to $O(L_0-1)$ and other symmetries
$U(L_1/2)\times\cdots\times U(L_{M-1}/2)\times O(L_M)$
remain unbroken.
In the case of $m^2C(\alpha_0)<0$,
a phase transition occurs when ${\cal M}^2(\alpha_0,R^\ast)=0$.
The critical radius $R^\ast$ is given by
	\begin{equation}
		R^\ast
	\sim
		\sqrt{-\frac{C(\alpha_0)}{m^2}}.
	\label{Rstar:alpha0}
	\end{equation}
If ${\cal M}^2(\alpha_1,R)>0$ with $L_0=0$ and $L_1\neq 0$,
the symmetry $H$ with $L_0=0$ is unbroken,
while if ${\cal M}^2(\alpha_1,R)<0$,
the $U(L_1/2)$ symmetry is spontaneously broken to $U(L_1/2-1)$
and other symmetries remain unbroken
\footnote{
In the case of $L_0=\cdots=L_{M-1}=0$,
{\it i.e.}
$L_M=N$,
$U(L_1/2)$ and $U(L_1/2-1)$ should be replaced by
$O(L_M)$ and $O(L_M-2)$, respectively.
}.
In the case of $m^2C(\alpha_1)<0$,
a phase transition occurs when ${\cal M}^2(\alpha_1,R^\ast)=0$.
The critical radius $R^\ast$ is given by
	\begin{equation}
		R^\ast
	\sim
		\sqrt{-\frac{C(\alpha_1)}{m^2}}.
	\label{Rstar:alpha1}
	\end{equation}

It should be noticed that if $\alpha^2_l\gg \lambda$,
the symmetry $U(L_l/2)$
($O(L_M)$ for $l=M$,
{\it i.e.} $\alpha_l=1/2$)
cannot be broken for small $R$, because $C(\alpha_l)\sim\alpha_l^2>0$
and hence ${\cal M}^2(\alpha_l,R)$ is positive for small $R\ll 1/m$
irrespective of the sign of $m^2$.
\footnote{
We have assumed that $\lambda\ll 1$
in order for perturbation theory to work.}
Especially, there is no possibility to break the $O(L_M)$
symmetry for small $R\ll 1/m$
because of $\alpha^2_M=(1/2)^2\gg\lambda$.

We see that the increasing sequence (\ref{order:M:small:alpha})
for small $R$ is identical to the sequence (\ref{order:of:M:large:R})
for large $R$.
We may then expect that the sequence (\ref{order:M:small:alpha})
or (\ref{order:of:M:large:R}) still persists in a whole range of $R$.
If this is true,
the phase structure of the model is rather simple and there are
only two phases:
The one is the unbroken phase in which no spontaneous symmetry
breaking occurs.
The other is the broken phase in which the symmetry
$O(L_0)$
($U(L_1/2)$ if $L_0=0$ and $L_1\neq 0$ or
$O(L_M)$ if $L_0=\cdots=L_{M-1}=0$)
is spontaneously broken to 
$O(L_0-1)$
($U(L_1/2-1)$ or $O(L_M-2)$)
and the remaining symmetries are unbroken.
We can,
at least,
show that the sequence (\ref{order:M:small:alpha}) holds
in a whole range of $R$ in a class of models with
$\alpha^2_l\gg\lambda$ for all $l\neq 0$
because the quantum corrections may then be less important
in ${\cal M}^2(\alpha_l,R)$
(except for ${\cal M}^2(\alpha_0,R)$)
in a whole range of $R$.

In the above analysis,
we have assumed that
$\alpha_l>\alpha^\ast$
for all $l\neq 0$.
If some of $\alpha_l$'s are smaller than $\alpha^\ast$,
the phase structure would then become complicated.
To see this,
let us consider,
for instance,
a model with $0<\alpha_l<\alpha^\ast/2$
for $l=1,2,\cdots,K$ and $\alpha_l>\alpha^\ast$ for
$l=K+1,K+2,\cdots,M$.
The increasing sequence (\ref{order:of:M:large:R})
still holds for large $R$.
For small $R$,
the sequence (\ref{order:M:small:alpha}) does not,
however,
hold but we have
	\begin{eqnarray}
		\lefteqn{{\cal M}^2(\alpha_K,R)<{\cal M}^2
		(\alpha_{K-1},R)<\cdots
		<{\cal M}^2(\alpha_0,R)}
		\nonumber \\
		&&<{\cal M}^2(\alpha_{K+1},R)<{\cal M}^2
		(\alpha_{K+2},R)<\cdots
		<{\cal M}^2(\alpha_M,R).
	\label{order:M:special:alpha}
	\end{eqnarray}
Note that the order of the first ($K+1$)
${\cal M}^2(\alpha_l,R)$'s for $l=0,1,\cdots,K$ 
is reversed from that
of Eq.(\ref{order:M:small:alpha}).
Since the order of the ($K+1$) ${\cal M}^2(\alpha_l,R)$'s for
$l=0,1,\cdots,K$ is completely opposite between for large $R$
and for small $R$,
we expect to have models that some of ${\cal M}^2(\alpha_l,R)$,
other than ${\cal M}^2(\alpha_0,R)$ and ${\cal M}^2(\alpha_K,R)$,
could take the lowest negative values in some regions of $R$.
If so,
the models would have multi-critical radii $R_a^\ast$
($a=1,2,\cdots$) and various symmetry phases.
It would be of interest to study those models in detail
but we will not proceed further since the full analysis would
require numerical computations and since the difference
${\cal M}^2(\alpha_l,R)-{\cal M}^2(\alpha_m,R)$ for
$0\leq l$, $m\leq K$ is tiny,
so that the analysis in Appendix A will not be justified
without taking vertex corrections into account,
as pointed out later.

We have so far discussed general properties of the $O(N)$ $\phi^4$
model with arbitrary twisted boundary conditions.
It may be instructive to examine some examples of twisted
boundary conditions which possess typical features discussed above.

\subsection*{(1) $U={\bf 1}_N$}

In this case,
all the fields obey the periodic boundary condition and
it does not break the $O(N)$ symmetry,
{\it i.e.}
$G=H=O(N)$.
For $m^2>0$,
the $O(N)$ symmetry is unbroken in a whole range of $R$ at the
classical level and also at the quantum level.
For $m^2=-\mu^2<0$,
the $O(N)$ symmetry would be broken to $O(N-1)$ spontaneously
in a whole range of $R$ at the classical level but the symmetry
restoration occurs for
$R\leq R^\ast\sim \sqrt{(N+2)\lambda/(96\pi^2\mu^2)}$
by quantum effects.
The phenomena of this symmetry restoration for small $R$ is
essentially the same as that
at high temperature\cite{DolanJackiw,Weinberg}
because expressions may become identical by identifying $2\pi R$
with the inverse temperature $T^{-1}$
in the imaginary time formulation.
The phase diagram of this model is summarized in Fig.2.
\begin{figure}
\begin{center}
\begin{tabular}{ccc}
\multicolumn{1}{l}{{\footnotesize (1)~~$U={\bf 1}_N$}}&  & 
\multicolumn{1}{l}{{\footnotesize (2)~~$U=-{\bf 1}_N$}}\\
  &  &  \\
\setlength\unitlength{0.5mm}
\begin{picture}(120,70)
{\footnotesize
\put(5,5){\vector(0,1){60}}
  \put(5,35){\vector(1,0){110}}
\multiput(50,5)(0,2){15}{\line(0,1){1}}
\put(5,66){\makebox(0,0)[b]{$m^2$}}
\put(116,35){\makebox(0,0)[l]{$R|m|$}}
\put(50,50){\makebox(0,0)[l]{$O(N)$}}
\put(27.5,20){\makebox(0,0){$O(N)$}}
\put(82.5,20){\makebox(0,0){$O(N-1)$}}
\put(50,4){\makebox(0,0)[t]{$R^*|m|$}}
\put(4,35){\makebox(0,0)[r]{$0$}}
\put(62,1.5){\makebox(0,0)[l]{$\sim \sqrt{
\frac{(N+2)\lambda}{96\pi^2}}$}}
}
\end{picture}
\label{U=1}

&\ \ &
\setlength\unitlength{0.5mm}
\begin{picture}(120,70)
\put(5,5){\vector(0,1){60}}
  \put(5,35){\vector(1,0){110}}
\multiput(50,5)(0,2){15}{\line(0,1){1}}
{\footnotesize
\put(5,66){\makebox(0,0)[b]{$m^2$}}
\put(116,35){\makebox(0,0)[l]{$R|m|$}}
\put(50,50){\makebox(0,0)[l]{$O(N)$}}
\put(27.5,20){\makebox(0,0){$O(N)$}}
\put(82.5,20){\makebox(0,0){$O(N-2)$}}
\put(50,4){\makebox(0,0)[t]{$R^*|m|$}}
\put(62,0.1){\makebox(0,0)[l]{$\sim \frac{1}{2}$}}
\put(4,35){\makebox(0,0)[r]{$0$}}
}
\end{picture}

\end{tabular}
\caption{
{ \footnotesize
The phase diagrams of the $O(N)$ $\phi^4$ model with $U={\bf 1}_N$
and $-{\bf 1}_N$ are represented in $(1)$ and $(2)$, respectively.
A critical radius $R^*$ appears for $m^2<0$ and is approximately 
given by $\sqrt{(N+2)\lambda/(96\pi^2|m|^2)}$ for $U={\bf 1}_N$
and $1/(2|m|)$ for $U=-{\bf 1}_N$.
}}

\end{center}
\end{figure}
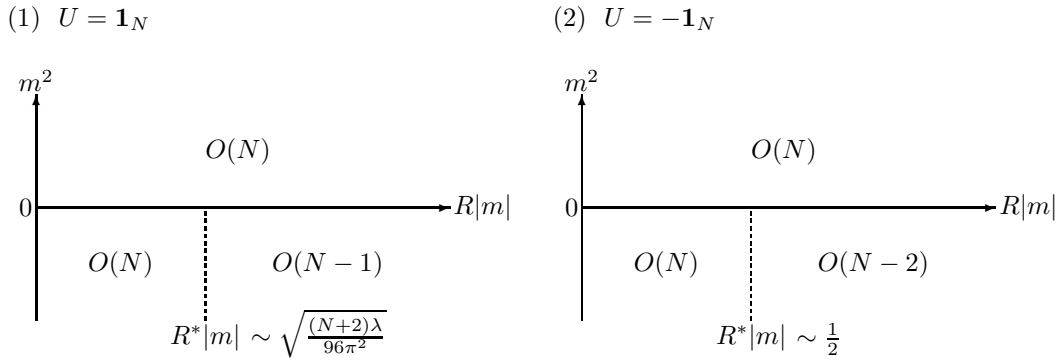


\subsection*{(2) $U=-{\bf 1}_N$}

In this case,
all the fields obey the antiperiodic boundary condition,
which does not break the $O(N)$ symmetry,
{\it i.e.}
$G=H=O(N)$.
The ``effective'' squared mass ${\cal M}^2(1/2,R)$ is given by
	\begin{equation}
		{\cal M}^2(1/2,R)
	=
		m^2+\left(\frac{1}{2R}\right)^2
		+\Delta m^2(1/2,R)\ .
	\label{M:all:fields:antiperiodic}
	\end{equation}
For large $R$,
the mass correction $\Delta m^2(1/2,R)$ is exponentially suppressed.
For small $R$,
$\Delta m^2(1/2,R)$ will be proportional to $\lambda/R^2$
but again less important compared to the second term on
the right-hand side of Eq.(\ref{M:all:fields:antiperiodic}),
as long as $\lambda\ll 1$.
In fact,
quantum corrections are irrelevant to determine the phase
structure in a whole range of $R$.
For $m^2>0$,
the $O(N)$ symmetry remains unbroken in a whole range of $R$,
while for $m^2=-\mu^2<0$,
the $O(N)$ symmetry is spontaneously broken to $O(N-2)$
(but not $O(N-1)$)
for $R>R^\ast\sim 1/(2\mu)$ and is restored for $R\leq R^\ast$.
It should be emphasized that the mechanism of this symmetry
restoration is different from the previous one of the model with
$U={\bf 1}_N$ and that the present symmetry restoration
has a classical origin.
This may be seen from the fact that $R^\ast$ is of order $1/\mu$,
but not $\sqrt{\lambda}/\mu$, in the present model.
The phase diagram of this model is summarized in Fig.2.
\begin{figure}
\begin{center}
\begin{tabular}{ccc}
\multicolumn{3}{l}{\footnotesize{
(3) ~$U=\left(
\begin{array}{cc} 
        {\bf 1}_{L_0} & 0 \\ 
0        & -{\bf 1}_{N-L_0} 
\end{array}
\right)$ }}\\
  &  &  \\
\setlength\unitlength{0.5mm}
\begin{picture}(120,70)
{\footnotesize
\put(5,5){\vector(0,1){60}}
  \put(5,35){\vector(1,0){110}}
\multiput(50,35)(0,2){15}{\line(0,1){1}}
\put(5,66){\makebox(0,0)[b]{$m^2$}}
\put(116,35){\makebox(0,0)[l]{$R|m|$}}

\put(22,56){\makebox(0,0){$O(L_0-1)$}}
\put(29,48){\makebox(0,0){$\times O(N-L_0)$}}

\put(70,56){\makebox(0,0){$O(L_0)$}}
\put(88,48){\makebox(0,0){$\times O(N-L_0)$}}
\put(60,20){\makebox(0,0){$O(L_0-1) \times O(N-L_0)$}}

\put(63.5,70.5){\makebox(0,0)[l]{$\sim \sqrt{\frac{(N-3L_0-4)\lambda }{
192{\pi}^2}}$}}

\put(50,66.5){\makebox(0,0)[b]{$R^*|m|$}}

\put(4,35){\makebox(0,0)[r]{$0$}}
}

\end{picture}
\label{U=a}
&\ \ &
\setlength\unitlength{0.5mm}
\begin{picture}(120,70)

{\footnotesize 
\put(5,5){\vector(0,1){60}}
  \put(5,35){\vector(1,0){110}}
\multiput(50,5)(0,2){15}{\line(0,1){1}}
\put(5,66){\makebox(0,0)[b]{$m^2$}}
\put(116,35){\makebox(0,0)[l]{$R|m|$}}

\put(18,26){\makebox(0,0){$O(L_0)$}}
\put(29,18){\makebox(0,0){$\times O(N-L_0)$}}

\put(70,26){\makebox(0,0){$O(L_0-1)$}}
\put(88,18){\makebox(0,0){$\times O(N-L_0)$}}
\put(60,50){\makebox(0,0){$O(L_0) \times O(N-L_0)$}}

\put(50,3.5){\makebox(0,0)[t]{$R^*|m|$}}

\put(4,35){\makebox(0,0)[r]{$0$}}
\put(62,0.5){\makebox(0,0)[l]{$\sim \sqrt{\frac{-(N-3L_0-4)\lambda }
{192{\pi}^2}}$}}
}
\end{picture}
\\
 & &  \\ 
{\footnotesize (3-a)~~$0 < L_0 < (N-4)/3$}& &
{\footnotesize (3-b)~~$N > L_0 >(N-4)/3$}\\
\end{tabular}
\end{center}
\caption{{\footnotesize
The phase diagrams of the $O(N)$ $\phi^4$ model with the 
twist matrix $U$ is represented in (3-a) for $0 < L_0 < (N-4)/3$
and in (3-b) for $N > L_0 > (N-4)/3$.
A critical radius $R^*$ appears if $m^2>0$ and $0 < L_0 < (N-4)/3$
or if $m^2<0$ and $N > L_0 > (N-4)/3$,
and it is approximately given by 
$\sqrt{(N-3L_0-4)\lambda /(192{\pi}^2 m^2)}$.
}} 

\end{figure}
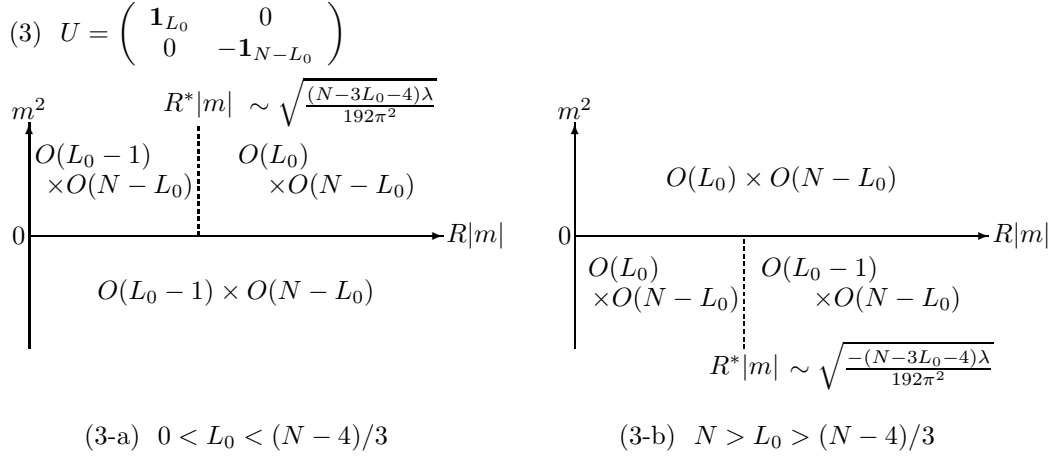
\subsection*{(3)
	$
		U
	=
		\left(
			\begin{array}{cc}
				{\bf 1}_{L_0}	&	0		\\
				0			&-{\bf 1}_{N-L_0}
			\end{array}
		\right)
	$
	}

Since the twist matrix $U$ is not proportional to the identity matrix,
the boundary condition (\ref{generalboundary}) explicitly breaks
the $O(N)$ symmetry down to $O(L_0)\times O(N-L_0)$,
which is the subgroup of $O(N)$ commuting with $U$.

For $m^2>0$,
the $O(L_0)\times O(N-L_0)$ symmetry is unbroken in a
whole range of $R$ if $N>L_0>(N-4)/3$,
but is broken to $O(L_0-1)\times O(N-L_0)$ for $R<R^\ast$
if $0<L_0<(N-4)/3$,
in spite of {\it positive} $m^2$.
The critical radius $R^\ast$ is given by
$R^\ast\sim \sqrt{(N-3L_0-4)\lambda/(192\pi^2m^2)}$.
This symmetry breaking for small $R$ seems strange
from the analogy with high temperature behavior of
scalar field theories
\footnote{
Weinberg has found a model in which the symmetry
breaking occurs at high temperature\cite{Weinberg},
but his model is assumed to have a negative coupling,
which is the origin of the symmetry breaking at high temperature.
On the other hand,
our model does not have any negative coupling.}.

For $m^2=-\mu^2<0$,
the $O(L_0)\times O(N-L_0)$ symmetry is broken to
$O(L_0-1)\times O(N-L_0)$ in a whole range of
$R$ if $0<L_0<(N-4)/3$,
but is restored for $R\leq R^\ast$ if $N>L_0>(N-4)/3$.
The mechanism of this symmetry restoration is essentially
the same as that found in the model with $U={\bf 1}_N$.
The phase diagram of this model is summarized in Fig.3.
\vskip5mm
We would like to finally discuss vertex corrections and higher
order ones,
which have been ignored in the above analysis.
Radiative corrections would induce the following type of vertex
corrections:
$
	\Delta\lambda_{lm}(R)
	(\phi_{a_l}^{(\alpha_l)})^2
	(\phi_{a_m}^{(\alpha_m)})^2
$.
Since quantum corrections do not respect the $O(N)$ symmetry,
the vertex correction $\Delta\lambda_{lm}(R)$ will,
in general,
depend on $l$ and $m$,
and is found to be of order $\lambda^2/(Rm)$ for small $R$.
\footnote{
For models with $L_0=0$,
$\Delta\lambda_{lm}(R)$ will be order
$\lambda^2\ln R$ and less important.}
Thus,
perturbation theory would be broken down at $R\sim \lambda/m$
because the vertex corrections would become the same order of
$\lambda$ at $R\sim\lambda/m$.
Fortunately,
the critical radii $R^\ast$ found in our analysis are
of order $\sqrt{\lambda}/|m|$ or $1/|m|$,
so that the phase transitions at $R=R^\ast$ can safely
be concluded to indeed occur.

One might still doubt the conclusion given in Appendix A
because vertex corrections have not been taken into account there.
We can,
however,
show that the inclusion of vertex corrections
does not change the phase
structure,
as long as
	\begin{equation}
		\frac{{\cal M}^2(\alpha_l,R)-{\cal M}^2
		(\alpha_P,R)}{{\cal M}^2(\alpha_P,R)}
	\gg
		\frac{\Delta\lambda}{\lambda}.
	\end{equation}
In fact,
the inclusion of vertex corrections does not modify the values
of the critical radii $R^\ast$ (if any)
but merely changes the non-vanishing vacuum expectation 
values slightly.

One might also worry about higher order corrections,
which could spoil the one-loop analysis for small $R$.
Weinberg has argued that the leading contribution at high
temperature will come from one-loop corrections
in perturbation theory\cite{Weinberg}.
This argument may be applied to our problem
and the qualitative features of our one-loop results will be trusted.

\section{QUANTUM EFFECTS IN $D>4$}\label{D>4}

In the previous section, we have investigated the vacuum structure
of the $O(N)$ $\phi^4$ model in $D=4$ dimensions.
In this section, 
we shall briefly discuss general features of the model in $D>4$  
dimensions.

Since the coupling constant $\lambda$ has the mass dimension
$-(D-4)$ for $D>4$, we
need to specify the mass scale of $\lambda$.
In the following analysis, we will assume 
that $|m|\lambda^{\frac{1}{D-4}}\ll1$. 
This relation may naturally be understood from an effective 
theory point of view, in which the scale of $\lambda$ will be taken
to be on the order of a cutoff of the theory\footnote{
Since the $O(N)$ $\phi^4$ model is not renormalizable
for $D>4$, the model should be understood as an effective theory with 
an ultraviolet cutoff.}. 
Then, the mass $|m|$ should be much below the cutoff, otherwise the
particle with the mass $|m|$ would decouple at low energies.

Since they are two mass scale of $|m|$ and $\lambda^{-\frac{1}{D-4}}$  
with the relation $|m|\ll\lambda^{-\frac{1}{D-4}}$,
it may be convenient to discuss the phase structure for
the following three regions separately:
(i) $R >_{{}_{{}_{\!\!\!\!\!\!\!{\textstyle \sim}}}} |m|^{-1}$,
(ii) $|m|^{-1} >_{{}_{{}_{\!\!\!\!\!\!\!{\textstyle \sim}}}}
R >_{{}_{{}_{\!\!\!\!\!\!\!{\textstyle \sim}}}}
\lambda^{\frac{1}{D-4}}$
and
(iii) $R <_{{}_{{}_{\!\!\!\!\!\!\!{\textstyle \sim}}}}
\lambda^{\frac{1}{D-4}}$.
In the region (i) 
$(R >_{{}_{{}_{\!\!\!\!\!\!\!{\textstyle \sim}}}} |m|^{-1})$, 
radiative corrections will be 
irrelevant because they are suppressed exponentially. Thus, the
phase structure should be the same as the classical one. 
In the region (ii) $(|m|^{-1} 
>_{{}_{{}_{\!\!\!\!\!\!\!{\textstyle \sim}}}} R 
>_{{}_{{}_{\!\!\!\!\!\!\!{\textstyle \sim}}}} 
\lambda^{\frac{1}{D-4}})$,
the one-loop mass corrections $\Delta m^2(\alpha_l,R)$ will 
approximately be given by (see the Appendix B)
\begin{equation}
	\Delta m^2(\alpha_l,R)\sim
	\lambda\frac{\overline{C}(\alpha_l)}{R^{D-2}},
\end{equation}
where 
\begin{equation}
	\overline{C}(\alpha_l)=
		\frac{\Gamma(\frac{D}{2}-1)}{2^D\pi^{\frac{3}{2}D-2}}
		\sum_{m=0}^M (L_m+2 \delta_{m,l})
		\sum_{n=1}^\infty \frac{\cos(2n\pi \alpha_m)}
	{n^{D-2}}.
\end{equation}
Then, we see that
$\Delta m^2(\alpha_0,R)$
becomes important in ${\cal M}^2(\alpha_0,R)$,
while $\Delta m^2(\alpha_k,R)$ for $k\neq 0$ may be
less important compared to $\alpha^2_k / R^2$
in ${\cal M}^2(\alpha_k,R)$
(unless $\alpha_k^2\ll1$).
\begin{eqnarray}
	{\cal M}^2(\alpha_0,R) & \sim & m^2
		+\lambda
		\frac{{\overline C}(\alpha_0)}{R^{D-2}},\\
		\nonumber
	{\cal M}^2(\alpha_k,R) & \sim & m^2
		+
		\frac{\alpha_k^2}{R^2}\qquad {\rm for}\ \ k\neq 0.
\end{eqnarray}
It turns out that two types of phase transitions can occur and
they essentially have the same origin found in the previous section.
If $L_0\ne0$ and $m^2\overline{C}(\alpha_0)<0$, a phase transition occur at
\begin{equation}
R=R^*\sim\left( -\lambda\frac{\overline{C}(\alpha_0)}
{m^2}\right)^\frac{1}{D-2}.
\end{equation}
In the case of $m^2>0$ ($m^2<0$),
the symmetry $O(L_0)$ ($O(L_0-1)$) will be broken to $O(L_0-1)$
(restored to $O(L_0)$) for $R<R^*$.
Another type of phase transitions can occur for the models with
$L_0=0$ and $m^2<0$ at 
\begin{equation}
R=R^*\sim\frac{\alpha_1}{|m|}.
\end{equation}
The broken symmetry $U(L_1/2-1)$
$(O(L_1-2)$ if $\alpha_1=1/2)$
will be restored to $U(L_1/2)$ $(O(L_1))$
for $R<R^*$. It is easy to see that both of the critical radii
lie in the region (ii).

In the region (iii) 
$(R <_{{}_{{}_{\!\!\!\!\!\!\!{\textstyle \sim}}}} 
\lambda^{\frac{1}{D-4}})$,
one might expect that the classical results could drastically be 
changed 
because the one-loop mass corrections would become large compared
to $m^2$ and $\alpha^2_l/R^2$ in ${\cal M}^2(\alpha_l,R)$.
The region of 
$R <_{{}_{{}_{\!\!\!\!\!\!\!{\textstyle \sim}}}}
\lambda^{\frac{1}{D-4}}$ is,
however, outside the validity of perturbation theory and hence
we cannot draw any reliable conclusions 
from one-loop computations in this region.
To see this,
we first note that one-loop vertex corrections will be of order 
$\lambda^2/R^{D-4}$ for small $R$.
It then implies that perturbation theory would be broken down for 
$R <_{{}_{{}_{\!\!\!\!\!\!\!{\textstyle \sim}}}}
\lambda^{\frac{1}{D-4}}$ 
because the vertex corrections
would become larger than $\lambda$.
Another argument of the breakdown of perturbation theory for
$R <_{{}_{{}_{\!\!\!\!\!\!\!{\textstyle \sim}}}}
\lambda^{\frac{1}{D-4}}$ 
may be given as follows:
From an effective theory point of view, 
it is natural to take the scale of
$\lambda$ to be a cutoff of the theory, {\it i.e.} 
$\lambda\sim$(cutoff)$^{-(D-4)}$.
Then, all the mass scales,
including $R^{-1}$,
should be taken to be smaller than the cutoff or 
$\lambda^{-\frac{1}{D-4}}$.
This implies that 
$R > \lambda^{\frac{1}{D-4}}$.
Therefore, to analyze the phase structure for 
$R <_{{}_{{}_{\!\!\!\!\!\!\!{\textstyle \sim}}}}
\lambda^{\frac{1}{D-4}}$,
we need to take higher order corrections into account, 
but the analysis is beyond the scope of this paper.

\section{REANALYSIS WITH KALUZA-KLEIN MODES}\label{kaluza}

In this section,
we would like to reanalyze the $O(N)$ $\phi^4$ model on
$M^{D-1}\otimes S^1$ from
a viewpoint of the ($D-1$)-dimensional theory.
To this end,
let us start with the functional
	\begin{eqnarray}
			{\cal E}[\phi,R]
		&=&
			\int^{2\pi R}_0dy
			\left\{
				\frac{1}{2}\sum^M_{l=0}\sum^{L_l}_{a_l=1}
				\phi^{(\alpha_l)}_{a_l}
				\left[
					-\partial_y^2+m^2+\Delta m^2(\alpha_l,R)
				\right]
				\phi^{(\alpha_l)}_{a_l}\right.
		\nonumber
		\\
		& &
		\left.
		+
		\frac{\lambda}{8}
		\left(
			\sum_{l=0}^M\sum_{a_l=1}^{L_l}
			\left(\phi_{a_l}^{(\alpha_l)}\right)^2
		\right)^2
		\right\}\ ,
	\label{energy:kk}
	\end{eqnarray}
which corresponds to the potential term in a $(D-1)$-dimensional
point of view.
To avoid inessential complexities,
we may restrict our considerations to the case of $L_0,L_M={\rm even}$.
Then,
it may be convenient to introduce the $N/2$ complex fields which can
be expanded in the Fourier-series as
	\begin{equation}
		\frac{1}{\sqrt{2}}
		\left(
			\phi_{2b_l-1}^{(\alpha_l)}(x^\nu,y)
			+
			i\phi_{2b_l}^{(\alpha_l)}(x^\nu,y)
		\right)
	=
		\sum^\infty_{n=-\infty}
		\varphi_{b_l,n}^{(\alpha_l)}(x^\nu)
		\exp \left(i\frac{n+\alpha_l}{R}y\right)
	\label{fourier:series:phi}
	\end{equation}
for $l=0,1,\cdots,M$ and $b_l=1,2,\cdots,L_l/2$.

The analysis in the previous sections suggests that in order to
examine the vacuum structure it is sufficient to keep only the
lowest modes in Eq.(\ref{fourier:series:phi}).
Inserting Eq.(\ref{fourier:series:phi}) into Eq.(\ref{energy:kk})
and keeping only the lowest modes,
we obtain
	\begin{eqnarray}
		\lefteqn{
		\frac{1}{2\pi R}{\cal E}[\varphi,R]_{{\rm lowest\ mode}}
		}\nonumber\\
	&& =
		\sum^{M-1}_{l=0}\sum^{L_{l}/2}_{b_l=1}
		{\cal M}^2(\alpha_l,R)\big|\varphi^{(\alpha_l)}_{b_l,0}
		\big|^2
	+
		\sum_{b_M=1}^{L_{M}/2}{\cal M}^2(\alpha_M,R)
		\left(
			\big|\varphi_{b_M,0}^{(\alpha_M)}\big|^2
			+
			\big|\varphi_{b_M,-1}^{(\alpha_M)}\big|^2
		\right)
	\nonumber
	\\
	&&
	+
	\frac{\lambda}{2}
		\left(
				\sum^{M-1}_{l=0}\sum^{L_{l}/2}_{b_l=1}
					\big|\varphi_{b_l,0}^{(\alpha_l)}\big|^2
		+
				\sum_{b_M=1}^{L_{M}/2}
					\left(
						\big|\varphi^{(\alpha_M)}_{b_M,0}\big|^2
						+
						\big|\varphi^{(\alpha_M)}_{b_M,-1}\big|^2
					\right)
		\right)^2
	+
	\lambda
		\Bigg|\sum_{b_M=1}^{L_{M}/2}
			\varphi_{b_M,0}^{(\alpha_M)\ast}
			\varphi_{b_M,-1}^{(\alpha_M)}
		\Bigg|^2.
	\label{enegy:lowest:mode}
	\end{eqnarray}
We should make two comments here.
The first comment is that the squared masses of
the lowest modes $\varphi_{b_l,0}^{(\alpha_l)}$ are
just given by ${\cal M}^2(\alpha_l,R)$,
which appear in the ``effective'' potential $\bar{V}(\bar{\phi})$
in Eq.(\ref{vbar:quantum}).
The second comment is that for $l=M$
({\it i.e.} $\alpha_l=1/2$)
the modes $\varphi^{(\alpha_M)}_{b_M,0}$
and $\varphi_{b_M,-1}^{(\alpha_M)}$
are doubly degenerate.
This fact will explain why the 
symmetry $O(L_M)$ is broken to $O(L_M-2)$
but not $O(L_M-1)$,
as we will see below.

Let ${\cal M}^2(\alpha_P,R)$ be the lowest value of 
${\cal M}^2(\alpha_l,R)$'s.
Then,
it is not difficult to show that if ${\cal M}^2(\alpha_P,R)$ is 
positive,
all the vacuum expectation values of the lowest modes should vanish.
On the other hand,
if ${\cal M}^2(\alpha_P,R)$ is negative with $P\neq M$
({\it i.e.} $\alpha_P\neq 1/2$),
the vacuum configuration of the lowest modes is taken,
without loss of generality,
into the form
	\begin{equation}
		\varphi_{b_P,0}^{(\alpha_P)}
	=
		\delta_{b_P,1}
		\sqrt{-\frac{{\cal M}^2(\alpha_P,R)}{\lambda}}
	\label{vev:varphi}
	\end{equation}
and other modes should vanish.
This implies that the symmetry $U(L_P/2)$
($O(L_0)$ if $P=0$)
is spontaneously broken to $U(L_P/2-1)$
($O(L_0-1)$).
For $P=M$
({\it i.e.} $\alpha_P=1/2$),
we must be careful in determining how the
$O(L_M)$ symmetry is broken.
In this case,
only $\varphi_{b_M,0}^{(\alpha_M)}$ and
$\varphi_{b_M,-1}^{(\alpha_M)}$
could acquire non-vanishing vacuum expectation values
according to the conditions
	\begin{eqnarray}
		\sum^{L_{M}/2}_{b_M=1}
		\left(
			\big|\varphi_{b_M,0}^{(\alpha_M)}\big|^2
			+
			\big|\varphi_{b_M,-1}^{(\alpha_M)}\big|^2
		\right)
	&=&
		-\frac{{\cal M}^2(\alpha_M,R)}{\lambda},
	\nonumber
	\\
	\sum_{b_M=1}^{L_{M}/2}
		\varphi_{b_M,0}^{(\alpha_M)\ast}
		\varphi_{b_M,-1}^{(\alpha_M)}
	&=&
		0.
	\label{varphi:condition}
	\end{eqnarray}
Because the $O(L_M)$ symmetry is not manifest in terms of
the complex variables,
we may return to the original real fields $\phi_{a_M}^{(\alpha_M)}$
($a_M=1,2,\cdots,L_M$)
and expand them into the Fourier-series
with real coefficients as follows:
	\begin{equation}
		\phi_{a_M}^{(\alpha_M)}(x^\nu,y)
	=
		\sum_{n=0}^\infty
			\left\{
				A_{a_M,n}(x^\nu)
					\cos\left(\frac{n+1/2}{R}y\right)
			+
				B_{a_M,n}(x^\nu)
					\sin\left(\frac{n+1/2}{R}y\right)
			\right\}
	\label{fourier:sin:cos}
	\end{equation}
for $a_M=1,2,\cdots,L_M$.
The relations between $\{A_{a_M,0},B_{a_M,0}\}$
and $\{\varphi_{b_M,0}^{(\alpha_M)},\varphi_{b_M,-1}^{(\alpha_M)}\}$
are found to be
	\begin{eqnarray}
		\varphi_{b_M,0}^{(\alpha_M)}
	&=&
		\frac{1}{2\sqrt{2}}
		\left\{
			A_{2b_M-1,0}
			+
			B_{2b_M,0}
			+
			i\left(
				A_{2b_M,0}-B_{2b_M-1,0}
			\right)
		\right\},
	\nonumber
	\\
		\varphi_{b_M,-1}^{(\alpha_M)}
	&=&
		\frac{1}{2\sqrt{2}}
		\left\{
			A_{2b_M-1,0}
			-
			B_{2b_M,0}
			+
			i\left(
				A_{2b_M,0}+B_{2b_M-1,0}
			\right)
		\right\}.
	\label{varphi:A:B}
	\end{eqnarray}
Inserting Eqs.(\ref{varphi:A:B}) into Eqs.(\ref{varphi:condition}),
we find
	\begin{eqnarray}
		\sum_{a_M=1}^{L_M}
		\left(
			\left(A_{a_M,0}\right)^2
			+
			\left(B_{a_M,0}\right)^2
		\right)
	&=&
		-\frac{4{\cal M}^2(\alpha_M,R)}{\lambda},
	\nonumber
	\\
		\sum_{a_M=1}^{L_M}\left(A_{a_M,0}\right)^2
	&=&
		\sum^{L_M}_{a_M=1}\left(B_{a_M,0}\right)^2,
	\label{condition:A:B}
	\\
	\sum_{a_M=1}^{L_M}A_{a_M,0}B_{a_M,0}
	&=&
	0.
	\nonumber
	\end{eqnarray}
The first two conditions in Eqs.(\ref{condition:A:B}) require that
some of $A_{a_M,0}$ and $B_{a_M,0}$ must be non-vanishing.
By an appropriate $O(L_M)$ rotation,
we can put $A_{a_M,0}$ to be of the form
$A_{a_M,0}=\rho \delta_{a_M,1}$,
where $\rho=\sqrt{-2{\cal M}^2(\alpha_M,R)/\lambda}$.
Then,
by an appropriate $O(L_M-1)$ rotation which leaves
$A_{a_M,0}=\rho\delta_{a_M,1}$ invariant,
we can put $B_{a_M,0}$ to be of the form
$B_{a_M,0}=\rho(\sin\theta\delta_{a_M,1}+\cos\theta\delta_{a_M,2})$.
The third condition of Eqs.(\ref{condition:A:B}),
however,
requires $\sin\theta=0$.
Thus,
we have arrived at the conclusion that the vacuum configuration can,
without loss of generality,
be taken to be of the form $A_{a_M,0}=\rho\delta_{a_M,1}$ and
$B_{a_M,0}=\rho\delta_{a_M,2}$.
Noting that both $A_{a_M,0}$ and $B_{a_M,0}$ belong to the vector
representation of $O(L_M)$,
we conclude that the $O(L_M)$
symmetry is spontaneously broken to $O(L_M-2)$,
as it should be.
Therefore,
all the results derived above are consistent with those given in the
previous sections.

Let us next examine the mass spectrum of the Kaluza-Klein modes.
Inserting the Fourier expansions (\ref{fourier:series:phi})
into Eq.(\ref{energy:kk}),
we have
	\begin{eqnarray}
		{\cal E}[\varphi,R]
		&=&
		2\pi R
		\sum_{l=0}^M\sum_{b_l=1}^{L_{l}/2}\sum^\infty_{n=-\infty}
			{\cal M}^2_n(\alpha_l,R)\big|\varphi_{b_l,n}^{(\alpha_l)}\big|^2
		\nonumber
		\\
		&&
		+\ \ \mbox{$\varphi^3$ and $\varphi^4$ terms,}
		\label{energy:fourier:varphi}
	\end{eqnarray}
where
	\begin{equation}
		{\cal M}^2_n(\alpha_l,R)
	\equiv
		m^2
		+
		\left(\frac{n+\alpha_l}{R}\right)^2
		+
		\Delta m^2(\alpha_l,R).
	\label{def:mn}
	\end{equation}
If ${\cal M}^2_0(\alpha_l,R)$ are positive for all $l$,
other squared masses are also positive.
Hence,
there will be no symmetry breaking and ${\cal M}^2_n(\alpha_l,R)$
give the mass spectrum of the Kaluza-Klein modes
$\varphi_{b_l,n}^{(\alpha_l)}$.
If some of ${\cal M}^2_0(\alpha_l,R)$ are negative,
the spontaneous symmetry breaking occurs.
Let ${\cal M}^2_0(\alpha_P,R)$ be the lowest (negative)
value of ${\cal M}^2_0(\alpha_l,R)$'s.
Then,
$\varphi_{b_l,n}^{(\alpha_l)}$ will acquire the following
vacuum expectation values:
	\begin{equation}
		\langle\varphi_{b_l,n}^{(\alpha_l)}\rangle
	=
		\delta_{l,P}\delta_{b_l,1}\delta_{n,0}
		\sqrt{-\frac{{\cal M}^2_0(\alpha_P,R)}{\lambda}}.
	\label{vev:varphi:alpha}
	\end{equation}
Replacing $\varphi^{(\alpha_l)}_{b_l,n}$ by
$\tilde{\varphi}_{b_l,n}^{(\alpha_l)}+
\langle\varphi^{(\alpha_l)}_{b_l,n}\rangle$
in ${\cal E}[\varphi,R]$,
we find
	\begin{eqnarray}
		{\cal E}[\tilde{\varphi}+\langle\varphi\rangle,R]
	&=&
		2\pi R \sum^\infty_{n=-\infty}
			\left\{
				\sum_{l=0}^M\sum_{b_l=1}^{L_{l}/2}
					\left[
						{\cal M}^2_n(\alpha_l,R)-
						{\cal M}^2_0(\alpha_P,R)
					\right]
				\big|\tilde{\varphi}_{b_l,n}^{(\alpha_l)}\big|^2
			\right.
	\nonumber
	\\
	&&
	-{\cal M}^2_0(\alpha_P,R)
	\left.
	\left(
		\big|\tilde{\varphi}_{1,n}^{(\alpha_P)}\big|^2
		+
		\frac{1}{2}\tilde{\varphi}^{(\alpha_P)}_{1,-n}
			\tilde{\varphi}^{(\alpha_P)}_{1,n}
		+
		\frac{1}{2}\tilde{\varphi}_{1,-n}^{(\alpha_P)\ast}
			\tilde{\varphi}_{1,n}^{(\alpha_P)\ast}
		\right)
		\right\}
		\nonumber
		\\
		&&
		+\mbox{\ \ $\tilde{\varphi}^3$ and $\tilde{\varphi}^4$ terms,}
	\label{energy:varphi:vev}
	\end{eqnarray}
where we have ignored irrelevant constants.
It is not difficult to show that the mass eigenvalues of 
the Kaluza-Klein modes are given as follows:
\begin{equation}
    \begin{array}{c|c}
    \mbox{K-K modes}     
	&    
	\mbox{$({\rm mass})^2$ eigenvalues} \\ \hline
    {\rm Re}\tilde{\varphi}_{1,0}^{(\alpha_P)}
    &
	-2{\cal M}^2_0(\alpha_P,R)\\
    {\rm Im}\tilde{\varphi}_{1,0}^{(\alpha_P)}
    &
	0\\
    \mbox{$\tilde{\varphi}_{1,n}^{(\alpha_P)}$
    and $ \tilde{\varphi}_{1,-n}^{(\alpha_P)}$ ($n\geq 1$)}
    &
	\ m^2_{n,\pm}\\
    \tilde{\varphi}^{(\alpha_P)}_{b_P,n}\ (b_P\neq 1)
    &
	\ {\cal M}_n^2(\alpha_P,R)-{\cal M}^2_0(\alpha_P,R)\\
    \tilde{\varphi}_{b_l,n}^{(\alpha_l)}\ (l\neq P)
    &
	\ {\cal M}^2_n(\alpha_l,R)-{\cal M}^2_0(\alpha_P,R)
    \end{array}
\end{equation}
where
	\begin{equation}
		m^2_{n,\pm}
	=
		-{\cal M}^2_0(\alpha_P,R)
		+
		\left(\frac{n}{R}\right)^2
		\pm
		\sqrt{
			4\left(\frac{n\alpha_P }{R^2}\right)^2
			+
			{\cal M}^4_0(\alpha_P,R)
		}.
	\end{equation}
Remembering that ${\cal M}^2_0(\alpha_P,R)$ is negative
and that ${\cal M}^2_0(\alpha_P,R)<{\cal M}^2_0(\alpha_l,R)$ 
for $l\neq P$,
we find that all the squared mass eigenvalues
are positive semi-definite,
as they should be.
For $P\neq M$ ({\it i.e.} $\alpha_P\neq 1/2$),
the $L_P-1$ massless modes,
${\rm Im}\tilde{\varphi}_{1,0}^{(\alpha_P)}$
and
$\tilde{\varphi}_{b_P,0}^{(\alpha_P)}$
($b_P=2,3,\cdots,L_P/2$),
appear and turn out to correspond to the Nambu-Goldstone modes
associated with the broken generators of $U(L_P/2)/U(L_P/2-1)$
($O(L_0)/O(L_0-1)$ if $P=0$).
If $P=M$ ({\it i.e.} $\alpha_P=1/2$),
the additional $L_M-2$ massless modes,
$\tilde{\varphi}_{b_M,-1}^{(\alpha_M)}$
($b_M=2,3,\cdots,L_{M/2}$),
appear and all the $2L_M-3$ massless modes turn out to form
the Nambu-Goldstone modes associated with the broken
generators of $O(L_M)/O(L_M-2)$.

\section{DISCUSSIONS}\label{discussions}

In this paper,we have discussed general features of
scalar field theories on $M^{D-1}\otimes S^1$ and especially 
studied the vacuum structure of the $O(N)$ $\phi^4$ model
on $M^{D-1}\otimes S^1$, 
which is in a class of models whose vacuum configurations
can be given in a rather simple form of Eq.(\ref{expect:phi:and:U}).
As discussed in Sec.\ref{general},
it will not,
in general,
easy to find vacuum configurations of field theories
on non-simply connected spaces because we must,
in general, minimize the total energy
(but not the potential alone)
with twisted boundary conditions.
It would be a challenging problem to classify vacuum configurations 
for general field theories on non-simply connected spaces.

A trivial extension of the models discussed in this paper is
to consider a higher dimensional non-simply connected space,
like $T^n$ ($n$-dimensional torus).
Another possible extension may be given as follows:
Twisted boundary conditions for the $S^1$ direction may physically
be interpreted as ``magnetic flux" passing through the circle $S^1$.
In this point of view,
one may say that our analysis has been made for the field theories
on $M^{D-1}\otimes S^1$ with a non-trivial background of magnetic
flux passing through the circle $S^1$.
This suggests that field theories on $S^n$
with non-trivial backgrounds of,
for instance,
magnetic monopoles or instantons could possess
interesting properties,
just like those found in this paper.

The introduction of fermions may be a straightforward exercise
but the introduction of gauge fields will yield a new feature.
The numbers $\alpha_l$ and $L_l$ appearing in the boundary
conditions (\ref{alpha0boundary})--(\ref{alphamboundary})
are free parameters in scalar field theories.
Some of them,
however,
become dynamical if gauge degrees of freedom are introduced.
A component of gauge fields in the compactified
direction can acquire a non-vanishing vacuum expectation value%
\cite{Hosotani}.
Then,
it can always be transformed into 
a vanishing vacuum expectation value
by a (generally singular) gauge transformation.
The effect of the transformation turns out to appear in boundary
conditions of charged fields,
that is,
the transformation twists boundary conditions of charged fields%
\cite{HiguchiParker}.
Thus, we could have gauge field theories in which the various
interesting phenomena observed in this paper are 
dynamically realized.

Although the field theories considered here are not a new type
of field theories,
most of the interesting properties found in this paper have
been overlooked so far.
Since our analysis is far from complete,
there should be many other uncovered properties.
It would be worth while proceeding to study higher dimensional
field theories more systematically and thoroughly.

\section*{ACKNOWLEDGMENTS}

~~We would like to thank to J. Arafune, C.S. Lim, M. Tachibana,
K. Takenaga and S. Tanimura for valuable discussions and 
useful comments.
This work was supported in part by JSPS Research Fellowship
for Young Scientists (H.H) and by Grant-In-Aid for Scientific
Research (No.12640275) from the Ministry of Education,
Science, and Culture, Japan (M.S).

\section*{APPENDIX A: MINIMIZATION OF THE POTENTIAL
 $\bar{V}(\bar{\phi})$}

In this Appendix, we minimize the following type of the potential
consisting of the $M+1$ real scalar fields:
 \begin{equation}
	\bar{V}(\bar{\phi})
 =
	\sum_{l=0}^M \frac{1}{2}
	{m_l}^2{\bar{\phi_l}}^2
 +
	\frac{\lambda}{8}
	\left( 
		\sum_{l=0}^M{\bar{\phi_l}}^2 
	\right)^2,
\label{A1}
                \end{equation}
where the squared masses 
${m_l}^2$
 have been arranged as
\begin{equation}
  {m_0}^2 <
             {m_1}^2 <
                        \cdots <
                                    {m_M}^2  .
\end{equation}
For our purpose, it is convenient to rewrite the potential
(\ref{A1}) into the form
\begin{equation}
   \bar{V}(\bar{\phi}) =
  \frac{\lambda}{8}
                       \left(\sum_{l=0}^M{\bar{\phi_l}}^2 
                    +
                       \frac{2}{\lambda}{m_0}^2 \right)^2
                    +
                       \sum_{l=0}^M \frac{1}{2}
                       \left({m_l}^2-{m_0}^2 \right)
                      {\bar{\phi_l}}^2    ,
\label{A3}   
\end{equation}
where we have dropped an irrelevant constant .
Since we have assumed that 
${m_l}^2-{m_0}^2 > 0$  for $l \neq 0$,
the second term on the right-hand side of Eq.(\ref{A3})
is positive semi-definite.
It is then obvious that the minimum of 
$\bar{V}(\bar{\phi})$
can be realized by the configuration that for 
${m_0}^2 \ge 0 $, $ \bar{\phi}_l=0$
for all $l$,
and that for ${m_0}^2<0$,
$\bar{\phi}_l^2=\delta_{l,0}$ $({-2{m_0}^2}/{\lambda})$.

\section*{APPENDIX B: COMPUTATIONS OF
ONE-LOOP MASS CORRECTIONS}

In this Appendix,
we compute the one-loop mass corrections of the $O(N)$ $\phi^4$
model on $M^{D-1}\otimes S^1$ and examine the asymptotic behavior.
To this end,
it is convenient to take the basis of the fields
$
\{\phi^{(\alpha_0)}_{a_0},
\Phi^{(\alpha_k)}_{b_k},
\phi_{a_M}^{(\alpha_M)}\}
$
which satisfy the boundary conditions
(\ref{alpha0boundary}), (\ref{PhiandE}) and
(\ref{alphamboundary}), respectively
\footnote{
The fields $\phi_{a_0}^{(\alpha_0)}$ and $\phi_{a_M}^{(\alpha_M)}$
are real but the fields $\Phi^{(\alpha_k)}_{b_k}$ are complex.
}.
In this basis,
the propagator of each field is simply given by
	\begin{equation}
		\frac{-i}{p^2+\left(\frac{n+\alpha_l}{R}\right)^2+m^2}
	\qquad
		\mbox{for $l=0,1,\cdots,M$,}
	\label{propagator}
	\end{equation}
and the loop integral on $M^{D-1}\otimes S^1$ is
	\begin{equation}
		\frac{1}{2\pi R}
		\sum^\infty_{n=-\infty}
		\int
		\frac{d^{D-1}p}{(2\pi)^{D-1}}.
	\label{sum:integral}
	\end{equation}
The contribution form the self-energy graph depicted in Fig.\ref{mass}
is then given by
(up to the vertex and the symmetry factors)
	\begin{equation}
		\zeta^D(\alpha,R)
	\equiv
		\frac{1}{2\pi R}\sum^\infty_{n=-\infty}
		\int\frac{d^{D-1}p}{(2\pi)^{D-1}}
		\frac{-i}{p^2+\left(\frac{n+\alpha}{R}\right)^2+m^2}.
	\label{def:zeta:D}
	\end{equation}
Taking account of the vertex and the symmetry factors,
we find that the one-loop mass correction to the field
$\phi_{a_l}^{(\alpha_l)}$ ($l=0,1,\cdots,M$)
is given by
	\begin{equation}
		\Delta m^2(\alpha_l,R)
	=
		\frac{\lambda}{2}
		\sum_{m=0}^M
		(L_m+2\delta_{m,l})
		\zeta^D(\alpha_m,R)
		+
		\delta m^2,
	\label{1:loop:mass}
	\end{equation}
where $\delta m^2$ denotes the mass counter term which will
be determined later.
Using the formulas
	\begin{equation}
		\frac{1}{A}
	=
		\int^\infty_0ds \ {\rm e}^{-As},
	\label{1:A}
	\end{equation}
	\begin{equation}
		\sum^\infty_{n=-\infty}
		\exp\left[-s\left(\frac{n+\alpha}{R}\right)^2\right]
	=
		R\sqrt{\frac{\pi}{s}}
		\sum_{n=-\infty}^\infty
		\exp\left[-\frac{(n\pi R)^2}{s}+i2n\pi\alpha\right],
	\label{poisson}
	\end{equation}
and carrying out the $p$-integration,
we may rewrite Eq.(\ref{def:zeta:D}) as
	\begin{equation}
		\zeta^D(\alpha,R)
	=
		\frac{1}{2^D\pi^{D/2}}\sum^\infty_{n=-\infty}
		\int^\infty_0ds\ s^{-{\scriptstyle \frac{D}{2}}}
		\ \exp\left[
		-\frac{(n\pi R)^2}{s}-m^2s+i2n\pi\alpha
		\right].
	\label{zeta:mass}
	\end{equation}
Here,
we determine the mass counter term $\delta m^2$ by
demanding the following renormalization condition:
	\begin{equation}
		\Delta m^2(\alpha_l,R)\big|_{R=\infty}
	=
		0.
	\end{equation}
This requirement turns out to be equivalent to replace
$\zeta^D(\alpha,R)$ by
	\begin{equation}
		\zeta^D_{\rm ren}(\alpha,R)
	\equiv
		\frac{1}{2^D\pi^{{\scriptstyle \frac{D}{2}}}}\sum_{n\neq 0}
		\int^\infty_0ds\ s^{-{\scriptstyle \frac{D}{2}}}
		\ \exp\left[
			-\frac{(n\pi R)^2}{s}
			-m^2s
			+i2n\pi\alpha
			\right]
	\label{def:zeta:ren}
	\end{equation}
and to drop $\delta m^2$ in Eq.(\ref{1:loop:mass}).
We can now carry out the $s$-integration by using the formula
	\begin{equation}
		\int^\infty_0ds\ s^{-\nu-1}
		{\rm e}^{-As-\frac{B}{s}}
	=
		2\left(\frac{A}{B}\right)^{\frac{\nu}{2}}
		K_\nu\left(2\sqrt{AB}\right),
	\label{Bessel}
	\end{equation}
where $K_\nu(z)$ is the modified Bessel function.
The result is
	\begin{equation}
		\zeta^D_{\rm ren}(\alpha,R)
	=
		\frac{m^{{\scriptstyle \frac{D}{2}}-1}}
		{2^{D-2}\pi^{D-1}R^{{\scriptstyle \frac{D}{2}}-1}}
		\sum_{n=1}^\infty
		\frac{\cos(2n\pi\alpha)K_{{\scriptstyle \frac{D}{2}}-1}
		(2n\pi Rm)
		}{
		n^{{\scriptstyle \frac{D}{2}}-1}
		}.
	\label{zeta:Bessel}
	\end{equation}

\begin{figure}
\begin{center}

 \setlength\unitlength{0.75mm}
  \begin{picture}(120,90)
  \put(5,5){\vector(0,1){80}}
  \put(5,45){\vector(1,0){110}}
  \multiput(5,45)(5.4,0){20}{\line(0,1){0.5}}
  \multiput(5,45)(27,0){5}{\line(0,1){1}}
  \put(116,45){\makebox(0,0)[l]{\small $R$  }}
  \put(4,45){\makebox(0,0)[r]{$0$}}

  \put(32,47){\makebox(0,0)[b]{\small $0.5$}}
  \put(59,47){\makebox(0,0)[b]{\small $1.0$}}
  \put(86,47){\makebox(0,0)[b]{\small $1.5$}}
  \put(113,47){\makebox(0,0)[b]{\small  $2.0$}}

  \qbezier[400](15.4,62.5)(16.8,49.7)(32,47.5)
  \qbezier[400](13.1,85)(13.5,75)(15.4,62.5)
  \qbezier[900](115,45)(40.1,45)(32,47.5)
  \qbezier[20](13.1,26.3)(15.5,40.5)(26.6,42.5)
  \qbezier[18](13.1,26.3)(11,15.9)(10.4,5)
  \qbezier[44](115,45)(37.4,45)(26.6,42.5)
  \put(5,86){\makebox(0,0)[b]{ $\zeta_{\rm ren}^{D=4}$ }}

\end{picture}
\end{center}
\caption{{\footnotesize The solid (dotted)
 line represents the curve of 
 $\zeta_{\rm ren}^{D=4}(0,R)$
$(\zeta_{\rm ren}^{D=4}(1/2,R))$ with $m=1$.
The asymptotic forms of the function 
$\zeta_{\rm ren}^{D=4}(\alpha,R)$ 
for $R \rightarrow 0$ and $R \rightarrow \infty$ are given 
in Eqs.(\ref{zeta:D:4:small:R}) and
(\ref{zeta:ren:large:R}), respectively.}}

\label{zeta}

\end{figure}
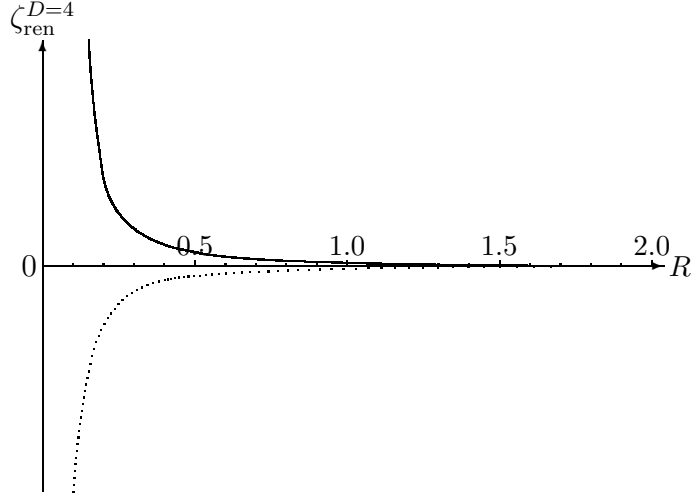

Let us next examine asymptotic behavior of the function
$\zeta^D_{\rm ren}(\alpha,R)$.
In the limit of $R\to 0$,
$\zeta^D_{\rm ren}(\alpha,R)$ reduces to
	\begin{equation}
		\zeta^D_{\rm ren}(\alpha,R)\ \ 
	\smash{\mathop{\hbox to 1.3cm{\rightarrowfill}}\limits^{R\to 0}}
		\ \ \frac{\rho^D(\alpha)}{R^{D-2}},
	\label{zeta:ren:small:R}
	\end{equation}
where
	\begin{equation}
		\rho^D(\alpha)
	=
		\frac{\Gamma({\textstyle \frac{D}{2}}-1)
				}{
				  2^{D-1}\pi^{{\scriptstyle \frac{3D}{2}}-2}
				  }
		\sum^\infty_{n=1}
		\frac{\cos(2n\pi\alpha)}{n^{D-2}}.
	\label{def:rho}
	\end{equation}
For even $D$,
the summation over $n$ can be carried out and the result is
	\begin{equation}
		\rho^D(\alpha)
	=
		\frac{
			(-1)^{{\scriptstyle \frac{D}{2}}}
			\Gamma({\textstyle \frac{D}{2}}-1)
			}{
			4(D-2)!\ \pi^{{\scriptstyle \frac{D}{2}}}
			}
		B_{D-2}(\alpha)
	\qquad
	\mbox{for $0\leq\alpha\leq 1$,}
	\label{rho:Bessel}
	\end{equation}
where $B_n(\alpha)$ denotes the Bernoulli polynomial.
Especially,
for the case of $D=4$,
we obtain
	\begin{equation}
		\zeta^{D=4}_{\rm ren}(\alpha,R)\ \ 
	\smash{\mathop{\hbox to 1.3cm{\rightarrowfill}}\limits^{R\to 0}}
		\ \ \frac{
			1-6\alpha+6\alpha^2
			}{
			48\pi^2R^2
			}.
	\label{zeta:D:4:small:R}
	\end{equation}
On the other hand,
in the limit of $R\to\infty$,
Eq.(\ref{zeta:Bessel}) reduces to
	\begin{equation}
		\zeta^D_{\rm ren}(\alpha,R)\ \ 
	\smash{
	\mathop{\hbox to 1.3cm{\rightarrowfill}}\limits^{R\to \infty}
	}
		\ \ \frac{
			\cos(2\pi\alpha)m^{{\scriptstyle \frac{D-3}{2}}}
			}{
			(2\pi)^{D-1}R^{{\scriptstyle \frac{D-1}{2}}}
			}\ 
		{\rm e}^{-2\pi Rm}.
	\label{zeta:ren:large:R}
	\end{equation}
The $R$ dependence of the function $\zeta^{D=4}_{\rm ren}(\alpha,R)$
for $\alpha=0$ and $1/2$ is schematically depicted in Fig.\ref{zeta}.
The solid (dotted) line represents the curve of the function
$\zeta^{D=4}_{\rm ren}(0,R)$
($\zeta^{D=4}_{\rm ren}(1/2,R)$).


\end{document}